\documentclass{article}%
\usepackage{amssymb}
\usepackage{cite}
\usepackage{amsmath}
\usepackage{makeidx}
\usepackage{amsfonts}
\usepackage{graphicx}%
\begin{document}

\author{Hartmut Wachter\thanks{E-Mail: Hartmut.Wachter@gmx.de}\\An der Schafscheuer 56\\D-91781 Wei\ss enburg, Federal Republic of Germany}
\title{Quantum Dynamics on the three-dimensional $q$-de\-formed Euclidean Space}
\maketitle
\date{}

\begin{abstract}
I extend the three-dimensional $q$-deformed Euclidean space by a time element
and discuss the algebraic structure of this quantum space together with its
differential calculi. Using the star-product formalism, I will give basic
operations of $q$-deformed analysis for the $q$-deformed Euclidean space with
a time element. I show that the time-evolution operator of a quantum system
living in the $q$-deformed Euclidean space is of the same form as in the
undeformed case. The reasonings also show that the well-known methods of
quantum dynamics apply to quantum systems living in the $q$-deformed Euclidean space.

\end{abstract}

\section{Introduction}

Several arguments say that space and time are discrete on a fundamental level
\cite{Garay:1995,Hagar:2014,Heisenberg:1930,Heisenberg:1938,Mead:1966zz}. To
answer the question of whether space and time are continuous or discrete, you
need to find out how the assumption of a discrete space-time changes physical
laws. The $q$-deformed Euclidean space provides a mathematical framework to
investigate this question \cite{Fichtmueller1996,Lorek:1997eh}.

In Ref.~\cite{Wachter:2019A}, I have begun to develop a formalism for the
quantum-theoretical description of free particles existing in the $q$-deformed
Euclidean space. In the present article, I will continue these considerations
and focus on the question of how to describe the time evolution of a
$q$-deformed quantum system. To this end, you need a $q$-deformed space that
has a time element. Considering a $q$-analog of Minkowski space could be one
way for this task \cite{CarowWatamura:1990nk, Lorek:1993tq, Lorek:1997eh}. The
$q$-deformed Minkowski space, however, has a very complex structure making
calculations very time-consuming. Therefore, I am going to choose a different
way in this article, i.~e., I extend the algebra of the $q$-deformed Euclidean
space by an element behaving like a commutative parameter.

Now, I am going to describe the contents of the various sections in more
detail. Chap.~\ref{KapQuaZeiEle} summarizes the algebraic basis for an
understanding of the $q$-deformed Euclidean space. In
Chap.~\ref{KapTimEleEucQuaSpa}, I show how to extend the algebra of the
$q$-deformed Euclidean space by an element that commutes with all coordinate
generators of the $q$-deformed Euclidean space and behaves like a scalar
concerning the Hopf algebra $\mathcal{U}_{q}(\operatorname*{su}\nolimits_{2}%
)$.\footnote{The Hopf algebra $\mathcal{U}_{q}(\operatorname*{su}%
\nolimits_{2})$ forms a $q$-analog of the angular momentum algebra. For this
reason, it describes the symmetry of the $q$-deformed Euclidean space
\cite{Lorek:1997eh}.} Due to its properties, the new element functions as a
time element. Furthermore, I will show how to extend the two differential
calculi of the $q$-deformed Euclidean space by a partial derivative for the
time element.

To formulate equations of motion on the $q$-deformed Euclidean space, we need
some tools of a $q$-deformed multidimensional analysis. For this purpose, we
associate the $q$-deformed Euclidean space with a commutative coordinate
algebra by using the star-product formalism. In doing so, you can calculate
formulas for star-products, $q$-derivatives, $q$-integrals as well as
$q$-translations and $q$-exponentials
\cite{Bauer:2003,Wachter:2004ExpA,Wachter:2004A,Wachter:2007A,Wachter:2002A}.
In Chap.~\ref{KapQAnaTim}, I will briefly explain the corresponding
considerations and apply them to the Euclidean quantum space supplemented by a
time element. Since the time element behaves like a commutative parameter and
is independent of the space coordinates, the following situation arises: The
formulas derived for the Euclidean quantum space in my previous work remain
valid, and the operations concerning the time element are of the same form as
in the undeformed case.

Chap.~\ref{KapZeiEntOpeN} shows you can perform time displacements in the
$q$-deformed Euclidean space in the same way as in the undeformed case.
Consequently, the time-evolution operator for a physical system existing in
the $q$-deformed Euclidean space is of the same form as in the undeformed
case. Due to this fact, we can apply the well-known methods for describing the
time development of a physical system to the $q$-deformed Euclidean space. In
Chap.~\ref{KapSchHeiBil}, you can see, in particular, that the Schr\"{o}dinger
equation and the Heisenberg equation of motion apply again.

\section{The three-dimensional $q$-deformed Euclidean
space\label{KapQuaZeiEle}}

The three-di\-men\-sio\-nal $q$-de\-formed Euclidean space $\mathbb{R}_{q}%
^{3}$ is a three-di\-men\-sio\-nal representation of the Drin\-feld-Jim\-bo
al\-ge\-bra $\mathcal{U}_{q}(\operatorname*{su}_{2})$ \cite{Kulish:1983md,
1985LMaPh..10...63J, Drinfeld:1985rx, Faddeev:1987ih}. The latter is a
deformation of the universal enveloping algebra of the Lie algebra
$\operatorname*{su}_{2}$. Accordingly, the algebra $\mathcal{U}_{q}%
(\operatorname*{su}_{2})$ has three generators $T^{+}$, $T^{-}$, and $T^{3}$
which satisfy the following relations \cite{Lorek:1993tq}:%
\begin{align}
q^{-1}\hspace{0.01in}T^{+}T^{-}-q\,T^{-}T^{+}  &  =T^{3},\nonumber\\
q^{\hspace{0.01in}2}\hspace{0.01in}T^{3}T^{+}-q^{-2}\hspace{0.01in}T^{+}T^{3}
&  =(q+q^{-1})\hspace{0.01in}T^{+},\nonumber\\
q^{\hspace{0.01in}2}\hspace{0.01in}T^{-}T^{3}-q^{-2}\hspace{0.01in}T^{3}T^{-}
&  =(q+q^{-1})\hspace{0.01in}T^{-}. \label{VerRel1UqSU2}%
\end{align}

The algebra $\mathcal{U}_{q}(\operatorname*{su}\nolimits_{2})$ has a
\textit{Casimir operator}, i.~e. an element that commutes with the generators
$T^{+}$, $T^{-}$, and $T^{3}$ \cite{Wess:math-ph9910013}:%
\begin{equation}
\vec{T}^{\hspace{0.01in}2}=q^{\hspace{0.01in}2}\lambda^{-2}\tau^{1/2}%
+\lambda^{-2}\tau^{-1/2}+\tau^{-1/2}\hspace{0.01in}T^{+}T^{-}-q\hspace
{0.01in}\lambda^{-2}\lambda_{+}. \label{TQuaCasUqSu2}%
\end{equation}
Note that we have introduced the element%
\begin{equation}
\tau=1-\lambda T^{3}%
\end{equation}
and the constants%
\begin{equation}
\lambda=q-q^{-1},\qquad\lambda_{+}=q+q^{-1}.
\end{equation}
The different representations of $\mathcal{U}_{q}(\operatorname*{su}%
\nolimits_{2})$ are distinguished by the eigenvalues of the Casimir operator
in Eq.$~$(\ref{TQuaCasUqSu2}) and the states of the same representation are
distinguished by the eigenvalues of the generator $T^{3}$. If we label the
eigenvalues of $\vec{T}^{2}$ with $j$ and those of $T^{3}$ with $m$ then we
have for each $j=0,\frac{1}{2},1,\ldots$ a $(2\hspace{0.01in}j+1)$%
-di\-men\-si\-onal irreducible representation with the states $\left\vert
\hspace{0.01in}j,m\right\rangle ,$ $m\in\left\{  -j,-j+1,\ldots,j-1,j\right\}
$. For the actions of the generators of $\mathcal{U}_{q}(\operatorname*{su}%
\nolimits_{2})$ on the states $\left\vert \hspace{0.01in}j,m\right\rangle $
applies \cite{Lorek:1993tq}:%
\begin{align}
\vec{T}^{\hspace{0.01in}2}\left\vert \hspace{0.01in}j,m\right\rangle  &
=[[\hspace{0.01in}j]]_{q^{-2}}[[\hspace{0.01in}j+1]]_{q^{2}}\left\vert
\hspace{0.01in}j,m\right\rangle ,\nonumber\\
T^{3}\left\vert \hspace{0.01in}j,m\right\rangle  &  =q^{-1}[[2\hspace
{0.01in}m]]_{q^{-2}}\left\vert \hspace{0.01in}j,m\right\rangle
,\label{DarUqSu2Anf}\\[0.08in]
T^{+}\left\vert \hspace{0.01in}j,m\right\rangle  &  =q^{-1}\sqrt
{[[\hspace{0.01in}j+m+1]]_{q^{-2}}[[\hspace{0.01in}j-m]]_{q^{2}}}\left\vert
\hspace{0.01in}j,m+1\right\rangle ,\nonumber\\
T^{-}\left\vert \hspace{0.01in}j,m\right\rangle  &  =q\sqrt{[[\hspace
{0.01in}j+m]]_{q^{-2}}[[\hspace{0.01in}j-m+1]]_{q^{2}}}\left\vert
\hspace{0.01in}j,m-1\right\rangle . \label{DarUqSu2End}%
\end{align}
The expressions above depend on the so-called antisymmetric $q$-numbers, which
are defined as follows:%
\begin{equation}
\lbrack\lbrack n]]_{q}=\frac{1-q^{n}}{1-q}.
\label{DefAntSymQZah2}%
\end{equation}

The states of the $\mathcal{U}_{q}(\operatorname*{su}_{2})$-triplet form the
coordinates of the three-di\-men\-si\-onal Euclidean quantum space
$\mathbb{R}_{q}^{3}$:%
\begin{equation}
X^{-}=\left\vert 1,-1\right\rangle ,\qquad X^{3}=\left\vert 1,0\right\rangle
,\qquad X^{+}=\left\vert 1,1\right\rangle . \label{IdeDreiKooiZus}%
\end{equation}
Thus, we obtain the actions of the $\mathcal{U}_{q}(\operatorname*{su}_{2}%
)$-generators on the coordinates $X^{+}$, $X^{3}$, and $X^{-}$ from the
identities in Eqs.~(\ref{DarUqSu2Anf}) and (\ref{DarUqSu2End}) if we choose
$j=1$ and take into account the identifications of Eq.~(\ref{IdeDreiKooiZus}):%
\begin{align}
T^{-}\triangleright X^{-}  &  =0,\nonumber\\
T^{-}\triangleright X^{3}  &  =q^{3/2}\lambda_{+}^{1/2}X^{-},\nonumber\\
T^{-}\triangleright X^{+}  &  =q^{1/2}\lambda_{+}^{1/2}X^{3}%
,\label{WirSUq(2)DreiEukQuaAnf}\\[0.08in]
T^{3}\triangleright X^{-}  &  =-q^{2}\lambda_{+}\hspace{0.01in}X^{-}%
,\nonumber\\
T^{3}\triangleright X^{3}  &  =0,\nonumber\\
T^{3}\triangleright X^{+}  &  =q^{-2}\lambda_{+}\hspace{0.01in}X^{+}%
,\\[0.08in]
T^{+}\triangleright X^{-}  &  =q^{-1/2}\lambda_{+}^{1/2}X^{3},\nonumber\\
T^{+}\triangleright X^{3}  &  =q^{-3/2}\lambda_{+}^{1/2}X^{+},\nonumber\\
T^{+}\triangleright X^{+}  &  =0. \label{WirSUq(2)DreiEukQuaEnd}%
\end{align}
In this respect, the quantum space $\mathbb{R}_{q}^{3}$ forms a left module of
the Hopf algebra $\mathcal{U}_{q}(\operatorname*{su}_{2})$.

The Drin\-feld-Jim\-bo al\-ge\-bra $\mathcal{U}_{q}(\operatorname*{su}_{2}%
)$\ is a Hopf algebra as well \cite{Klimyk:1997eb}. For this reason, it has a
mapping $\Delta:\mathcal{U}_{q}(\operatorname*{su}_{2})\rightarrow
\mathcal{U}_{q}(\operatorname*{su}_{2})\otimes\mathcal{U}_{q}%
(\operatorname*{su}_{2})$ called coproduct. On the generators of
$\mathcal{U}_{q}(\operatorname*{su}_{2})$, the co-pro\-duct reads as follows
\cite{Lorek:1995ph, Wess:math-ph9910013}:%
\begin{align}
\Delta(T^{+})  &  =T^{+}\otimes1+\tau^{1/2}\otimes T^{+},\nonumber\\
\Delta(T^{-})  &  =T^{-}\otimes1+\tau^{1/2}\otimes T^{-},\nonumber\\
\Delta(T^{3})  &  =T^{3}\otimes1+\tau\otimes T^{3}. \label{KopSUq(2)Gen}%
\end{align}
Being a Hopf algebra, $\mathcal{U}_{q}(\operatorname*{su}_{2})$ has an
antipode $S:\mathcal{U}_{q}(\operatorname*{su}_{2})\rightarrow$ $\mathcal{U}%
_{q}(\operatorname*{su}_{2})$ and a co-unit $\varepsilon:\mathcal{U}%
_{q}(\operatorname*{su}_{2})\rightarrow%
%TCIMACRO{\U{2102} }%
%BeginExpansion
\mathbb{C}
%EndExpansion
$ as well. For the antipode applies%
\begin{equation}
S(T^{\pm})=-\hspace{0.01in}\tau^{-1/2}\hspace{0.01in}T^{\pm},\qquad
S(T^{3})=-\hspace{0.01in}\tau^{-1}T^{3}. \label{HopfStrTPluTMin}%
\end{equation}
The co-units of the generators $T^{+}$, $T^{-}$, and $T^{3}$ are equal to
zero:%
\begin{equation}
\varepsilon(T^{+})=\varepsilon(T^{-})=\varepsilon(T^{3})=0.
\label{AntUqSu2Gen}%
\end{equation}

The coproduct of $\mathcal{U}_{q}(\operatorname*{su}_{2})$ determines the
commutation relations between the generators of the Hopf algebra
$\mathcal{U}_{q}(\operatorname*{su}_{2})$ and the coordinates of the Euclidean
quantum space $\mathbb{R}_{q}^{3}$ since we have%
\begin{equation}
h\hspace{0.01in}v=\left(  h_{\left(  1\right)  }\triangleright v\right)
h_{\left(  2\right)  }\quad\text{for}\quad h\in\mathcal{U}_{q}%
(\operatorname*{su}\nolimits_{2}),\quad v\in\mathbb{R}_{q}^{3}.
\label{ComRelHV}%
\end{equation}
Note that we have written the coproduct in the so-called Sweedler notation,
i.~e. $\Delta(h)=h_{\left(  1\right)  }\otimes h_{\left(  2\right)  }$. Taking
into account the actions in Eqs.~(\ref{WirSUq(2)DreiEukQuaAnf}%
)-(\ref{WirSUq(2)DreiEukQuaEnd}) as well as the expressions in
Eq.~(\ref{KopSUq(2)Gen}), the above identity leads to the following
commutation relations \cite{Lorek:1997eh}:%
\begin{align}
T^{-}X^{-}  &  =q^{2}X^{-}T^{-},\nonumber\\
T^{-}X^{3}  &  =X^{3}\hspace{0.01in}T^{-}+\hspace{0.01in}q^{3/2}\lambda
_{+}^{1/2}X^{-},\nonumber\\
T^{-}X^{+}  &  =q^{-2}X^{+}T^{-}+\hspace{0.01in}q^{1/2}\lambda_{+}^{1/2}%
X^{3},\label{VerRelUqSU2EukQuaDreAnf}\\[0.08in]
T^{3}X^{-}  &  =q^{4}X^{-}T^{3}-q^{2}\lambda_{+}\hspace{0.01in}X^{-}%
,\nonumber\\
T^{3}X^{3}  &  =X^{3}\hspace{0.01in}T^{3},\nonumber\\
T^{3}X^{+}  &  =q^{-4}X^{+}T^{3}+q^{-2}\lambda_{+}\hspace{0.01in}%
X^{+},\label{VerRelUqSU2EukQuaDreMit}\\[0.08in]
T^{+}X^{-}  &  =q^{2}X^{-}T^{+}+\hspace{0.01in}q^{-1/2}\lambda_{+}^{1/2}%
X^{3},\nonumber\\
T^{+}X^{3}  &  =X^{3}\hspace{0.01in}T^{+}+\hspace{0.01in}q^{-3/2}\lambda
_{+}^{1/2}X^{+},\nonumber\\
T^{+}X^{+}  &  =q^{-2}X^{+}T^{+}. \label{VerRelUqSU2EukQuaDreEnd}%
\end{align}

The three generators $X^{+}$, $X^{3}$, and $X^{-}$ of the quantum space
$\mathbb{R}_{q}^{3}$ are subject to the following commutation relations
\cite{Lorek:1997eh}:%
\begin{gather}
X^{3}X^{+}-q^{2}X^{+}X^{3}=0,\nonumber\\
X^{3}X^{-}-q^{-2}X^{-}X^{3}=0,\nonumber\\
X^{-}X^{+}-X^{+}X^{-}-(q-q^{-1})\hspace{0.01in}X^{3}X^{3}=0.
\label{RelQuaEukDre}%
\end{gather}
The relations above are completely determined by the requirement that they
have to be compatible with the commutations relations in
Eqs.~(\ref{VerRelUqSU2EukQuaDreAnf})-(\ref{VerRelUqSU2EukQuaDreEnd}). For
example, it must hold for all $A\in\{+,3,-\}$:%
\begin{equation}
T^{A}(X^{3}X^{+}-q^{2}X^{+}X^{3})=(X^{3}X^{+}-q^{2}X^{+}X^{3})\hspace
{0.01in}T^{A}.
\end{equation}
The space $\mathbb{R}_{q}^{3}\hspace{0.01in}\otimes\hspace{0.01in}%
\mathcal{U}_{q}(\operatorname*{su}_{2})$ together with the commutation
relations in Eqs.~(\ref{VerRel1UqSU2}), (\ref{VerRelUqSU2EukQuaDreAnf}%
)-(\ref{VerRelUqSU2EukQuaDreEnd}), and (\ref{RelQuaEukDre}) forms the
so-called \textit{left cross product algebra} $\mathbb{R}_{q}^{3}%
\rtimes\hspace{0.01in}\mathcal{U}_{q}(\operatorname*{su}\nolimits_{2})$
\cite{Majid:1996kd, Weixler:1993ph}.

The Euclidean quantum space $\mathbb{R}_{q}^{3}$ is also a $\ast$-al\-ge\-bra,
i.~e. it has a semilinear, involutive, and anti-multiplicative mapping. We call
this mapping \textit{quantum space conjugation}. If we indicate the conjugate
elements of a quantum space by a bar\footnote{A bar over a complex number
indicates complex conjugation.}, we can write the properties of the quantum
space conjugation as follows ($\alpha,\beta\in\mathbb{C}$ and $u,v\in
\mathbb{R}_{q}^{3}$):%
\begin{equation}
\overline{\alpha\,u+\beta\,v}=\overline{\alpha}\,\overline{u}+\overline{\beta
}\,\overline{v},\quad\overline{\overline{u}}=u,\quad\overline{u\,v}%
=\overline{v}\,\overline{u}.
\end{equation}
You can show that the conjugation of $\mathbb{R}_{q}^{3}$ respects the
commutation relations in Eq.~(\ref{RelQuaEukDre}) if the following applies
\cite{Lorek:1997eh}:%
\begin{equation}
\overline{X^{+}}=X_{+}=-\hspace{0.01in}q\hspace{0.01in}X^{-},\qquad
\overline{X^{3}}=X_{3}=X^{3},\qquad\overline{X^{-}}=X_{-}=-\hspace
{0.01in}q^{-1}X^{+}. \label{ConSpaKoo}%
\end{equation}

As we know, the Euclidean quantum space is a module of the Hopf algebra
$\mathcal{U}_{q}(\operatorname*{su}_{2})$. We require that the tensor product
of two Euclidean quantum spaces is also a module of $\mathcal{U}%
_{q}(\operatorname*{su}_{2})$. For this reason, we have to calculate the
action of $\mathcal{U}_{q}(\operatorname*{su}_{2})$ on a tensor product of two
Euclidean quantum spaces by using the co-pro\-duct of $\mathcal{U}%
_{q}(\operatorname*{su}_{2})$. Specifically, it applies to $h\in
\mathcal{U}_{q}(\operatorname*{su}_{2})$ and $u,v\in\mathbb{R}_{q}^{3}$
\cite{Klimyk:1997eb}:%
\begin{equation}
h\triangleright(u\otimes v)=h_{(1)}\triangleright u\otimes h_{(2)}%
\triangleright v.
\end{equation}

Due to the non-tri\-vial co-pro\-duct of $\mathcal{U}_{q}(\operatorname*{su}%
_{2})$, we cannot perform the multiplication on a tensor product
$\mathbb{R}_{q}^{3}\otimes\mathbb{R}_{q}^{3}$ by using the usual twist. But it
works if we use a so-called braiding map instead. In the case of two
coordinate generators this braiding map is given by the R-ma\-trix $\hat{R}%
{}{^{\hspace{0.01in}AB}}_{CD}$ of the three-di\-men\-si\-onal $q$-de\-formed
Euclidean space (together with a constant $k$):%
\begin{equation}
(1\otimes Y^{A})\cdot(X^{B}\otimes1)=k\hspace{0.02in}\hat{R}{}{^{\hspace
{0.01in}AB}}_{CD}\,X^{C}\otimes Y^{D}. \label{ZopRel}%
\end{equation}
Applying the action of the $\mathcal{U}_{q}(\operatorname*{su}_{2}%
)$-ge\-ne\-ra\-tors to the above equations results in a system of equations to
determine the entries of the R-ma\-trix of the Euclidean quantum space
\cite{Lorek:1993tq}.\footnote{The constant $k$ cannot be determined this way.}
The inverse matrix ${\hat{R}}^{-1}$ is a further solution of this system:%
\begin{equation}
(1\otimes Y^{A})\cdot(X^{B}\otimes1)=k^{-1}\hspace{0.02in}({\hat{R}}%
^{-1}{)^{AB}}_{CD}\,X^{C}\otimes Y^{D}. \label{InvZopRel}%
\end{equation}
In the following, we will refer to Eqs.~(\ref{ZopRel}) and (\ref{InvZopRel})
as \textit{braiding relations}.

The R-ma\-trix $\hat{R}{}{^{\hspace{0.01in}AB}}_{CD}$ for $\mathbb{R}_{q}^{3}$
is of block-diagonal form. The indices of its rows and columns are $++,$ $--,$
$+\hspace{0.01in}3,$ $3\hspace{0.01in}+,$ $3-,$ $-3,$ $+-,$ $33$, and $-+$. If
the upper indices refer to the rows and the lower ones to the columns of
$\hat{R}{}{^{\hspace{0.01in}AB}}_{CD}$, we can write the entries of the
non-vanishing blocks of this R-matrix as follows \cite{Lorek:1997eh}:%
\begin{gather}%
\begin{tabular}
[c]{ccc}
& $++$ & $--$\\\cline{2-3}%
$++$ & \multicolumn{1}{|c}{$1$} & $0$\\
$--$ & \multicolumn{1}{|c}{$0$} & $1$%
\end{tabular}
\ ,\label{RMatSOq3Anf}\\[0.03in]%
\begin{tabular}
[c]{ccc}
& $+\hspace{0.01in}3$ & $3\hspace{0.01in}+$\\\cline{2-3}%
$+\hspace{0.01in}3$ & \multicolumn{1}{|c}{$0$} & $q^{-2}$\\
$3\hspace{0.01in}+$ & \multicolumn{1}{|c}{$q^{-2}$} & $q^{-2}\lambda
\lambda_{+}$%
\end{tabular}
\ ,\quad%
\begin{tabular}
[c]{ccc}
& $3-$ & $-3$\\\cline{2-3}%
$3-$ & \multicolumn{1}{|c}{$0$} & $q^{-2}$\\
$-3$ & \multicolumn{1}{|c}{$q^{-2}$} & $q^{-2}\lambda\lambda_{+}$%
\end{tabular}
\ ,\\[0.03in]%
\begin{tabular}
[c]{llll}
& $+-$ & $33$ & $-+$\\\cline{2-4}%
$+-$ & \multicolumn{1}{|l}{$0$} & $0$ & $q^{-4}$\\
$33$ & \multicolumn{1}{|l}{$0$} & $q^{-2}$ & $q^{-3}\lambda\lambda_{+}$\\
$-+$ & \multicolumn{1}{|l}{$q^{-4}$} & $q^{-3}\lambda\lambda_{+}$ &
$q^{-3}\lambda^{2}\lambda_{+}$%
\end{tabular}
\ . \label{RMatSOq3End}%
\end{gather}

The R-ma\-trix of $\mathbb{R}_{q}^{3}$ has the eigenvalues $1$, $q^{-6}$, and
$-q^{-4}$, which correspond to three projectors $P_{S}$, $P_{T}$, and $P_{A}$.
For this reason, the R-matrix of $\mathcal{\mathbb{R}}_{q}^{3}$ shows the
following projector decomposition \cite{Lorek:1993tq}:%
\begin{equation}
\hat{R}=P_{S}+q^{-6}P_{T}-q^{-4}P_{A}.
\end{equation}
The projector $P_{A}$ is a $q$-analog of the antisymmetrizer, which maps on the
space of antisymmetric tensors of rank two. The projector $P_{S}$ is the
$q$-deformed trace-free symmetrizer, and $P_{T}$ is the $q$-de\-formed
trace-projector. With the help of the projector $P_{A}$, we regain the
commutation relations for the coordinates of $\mathcal{\mathbb{R}}_{q}^{3}$
[cf. Eq.~(\ref{RelQuaEukDre})]:%
\begin{equation}
(P_{A}){^{AB}}_{CD}\,X^{C}X^{D}=0. \label{ProASymKooRel}%
\end{equation}
The projector $P_{T}$ leads us to a $q$-analog of the Euclidean metric
\cite{Lorek:1997eh}. In this respect, it applies to the metric $g^{\hspace
{0.01in}AB}$ of the $q$-deformed Euclidean space and its inverse $g_{CD}$:%
\begin{equation}
(P_{T}){^{AB}}_{CD}=\frac{1}{g^{EF}g_{EF}}\,g^{AB}g_{CD}.
\end{equation}
The relation above implies (row and column indices have the order $+,3,-$):%
\begin{equation}
g_{AB}=g^{\hspace{0.01in}AB}=\left(
\begin{array}
[c]{ccc}%
0 & 0 & -\hspace{0.01in}q\\
0 & 1 & 0\\
-\hspace{0.01in}q^{-1} & 0 & 0
\end{array}
\right)  .
\end{equation}
Using the $q$-deformed Euclidean metric, we can raise and lower indices:%
\begin{equation}
X_{A}=g_{AB}X^{B},\quad X^{A}=g^{\hspace{0.01in}AB}X_{B}. \label{HebSenInd}%
\end{equation}

The projectors $P_{S}$ and $P_{T}$ determine the commutation relations between
the coordinate differentials d$X^{+}$, d$X^{3}$, and d$X^{-}$
\cite{CarowWatamura:1990zp,Wess:1990vh}:%
\begin{equation}
(P_{S}){^{AB}}_{CD}\,\text{d}X^{C}\text{d}X^{D}=0,\qquad(P_{T}){^{AB}}%
_{CD}\,\text{d}X^{C}\text{d}X^{D}=0. \label{RelKooDif}%
\end{equation}
The identities above imply the following commutation relations between the
spatial coordinates and their differentials \cite{Lorek:1995ph}:%
\begin{equation}
X^{A}\text{d}X^{B}=q^{4}\hat{R}{^{\hspace{0.01in}AB}}_{CD}\,\text{d}X^{C}%
X^{D}. \label{RelKooDifKoo}%
\end{equation}

\section{Time element for the quantum space $\mathbb{R}_{q}^{3}$%
\label{KapTimEleEucQuaSpa}}

\subsection{Commutation relations\label{KapComRel}}

In the following, we are describing how to extend the algebra of the quantum
space $\mathbb{R}_{q}^{3}$ by a time element $X^{0}$. We require that $X^{0}$
transforms under the action of the Hopf algebra $\mathcal{U}_{q}%
(\operatorname*{su}\nolimits_{2})$ like a scalar $\left\vert 0,0\right\rangle
$, i.~e.%
\begin{equation}
T^{A}\triangleright X^{0}=0\text{\qquad with\qquad}A\in\{+,3,-\}.
\label{SkaZeiTran}%
\end{equation}
Due to Eq.~(\ref{ComRelHV}) of the previous chapter, this implies that $X^{0}$
commutes with all generators of the Hopf algebra $\mathcal{U}_{q}%
(\operatorname*{su}\nolimits_{2})$:%
\begin{equation}
T^{A}X^{0}=X^{0}T^{A}.
\end{equation}
Additionally, we require that extending $\mathbb{R}_{q}^{3}$ by $X^{0}$ does
not change the commutation relations between the coordinate generators $X^{+}%
$, $X^{3}$, and $X^{-}$ [cf. Eq.~(\ref{RelQuaEukDre}) of the previous
chapter]. To achieve this, we assume that the commutation relations between
$X^{0}$ and the coordinate generators of the quantum space
$\mathcal{\mathbb{R}}_{q}^{3}$ are of the following form ($A\in\{+,3,-\}$):%
\begin{equation}
X^{0}X^{A}=c_{A}\hspace{0.01in}X^{A}X^{0}. \label{ZusRelExtDreEukQUa}%
\end{equation}
To determine the unknown coefficients $c_{A}$, we require that the commutation
relations between the space-time coordinates and the $\mathcal{U}%
_{q}(\operatorname*{su}\nolimits_{2})$-ge\-ne\-ra\-tors do not change the
relations in Eq.~(\ref{ZusRelExtDreEukQUa}):%
\begin{equation}
T^{B}\hspace{0.01in}(X^{0}X^{A}-c_{A}\hspace{0.01in}X^{A}X^{0})=(X^{0}%
X^{A}-c_{A}\hspace{0.01in}X^{A}X^{0})\hspace{0.01in}T^{B}.
\end{equation}
This requirement implies that the coefficients $c_{A}$ are equal to each
other:%
\begin{equation}
c=c_{+}=c_{3}=c_{-}.
\end{equation}
The value of the parameter $c$ follows from the assumption that the relations
in Eq.~(\ref{ZusRelExtDreEukQUa}) have to be invariant under quantum space
conjugation. Taking into account the conjugation properties of the time
element, i.~e.%
\begin{equation}
\overline{X^{0}}=X_{0}=X^{0}, \label{ConTim}%
\end{equation}
conjugating Eq.~(\ref{ZusRelExtDreEukQUa}) leads to the condition $\left\vert
c\hspace{0.01in}\right\vert ^{2}=1$. Thus, we end up with the following result:%
\begin{equation}
X^{0}X^{A}=X^{A}X^{0}. \label{VerRelZeiRauKooDreQua}%
\end{equation}

However, the condition $\left\vert c\hspace{0.01in}\right\vert ^{2}=1$ has the solution
$c=-1$, as well. This second solution is relevant if we extend the exterior
algebra $\Lambda_{q}(\mathbb{R}_{q}^{3})$ by the coordinate differential
d$X^{0}$. In analogy to Eq.~(\ref{SkaZeiTran}), we assume that d$X^{0}$
transforms under the action of $\mathcal{U}_{q}(\operatorname*{su}%
\nolimits_{2})$ like a scalar ($A\in\{+,3,-\}$):%
\begin{equation}
T^{A}\triangleright\text{d}X^{0}=0.
\end{equation}
Note that the relations in Eq.~(\ref{RelKooDif}) or Eq.~(\ref{RelKooDifKoo})
of the previous chapter define the commutation relations of the exterior
algebra $\Lambda_{q}(\mathbb{R}_{q}^{3})$. These relations should remain the
same if we introduce d$X^{0}$. Moreover, the coordinate differentials should
have the same conjugation properties as the corresponding coordinates:%
\begin{equation}
\overline{\text{d}X^{A}}=\text{d}X_{A}=g_{AB}\,\text{d}X^{B},\qquad
\overline{\text{d}X^{0}}=\text{d}X_{0}=\text{d}X^{0},
\end{equation}
These requirements lead to%
\begin{equation}
\text{d}X^{0}\text{d}X^{A}=c\,\text{d}X^{A}\text{d}X^{0}%
\end{equation}
with $\left\vert c\hspace{0.01in}\right\vert ^{2}=1$. Since d$X^{0}$ and d$X^{A}$ are
elements of an exterior algebra, we now choose $c=-1$. This way, we finally
get:%
\begin{equation}
\text{d}X^{0}\text{d}X^{A}=-\hspace{0.01in}\text{d}X^{A}\text{d}X^{0}%
,\quad\text{d}X^{0}\text{d}X^{0}=0. \label{VerDiffXAX0ExtEuk}%
\end{equation}

For the sake of completeness, we note that the relations in
Eq.~(\ref{VerDiffXAX0ExtEuk}) are compatible with the following commutation
relations:%
\begin{equation}
X^{0}\hspace{0.01in}\text{d}X^{A}=\hspace{0.01in}\text{d}X^{A}X^{0},\quad
X^{A}\hspace{0.01in}\text{d}X^{0}=\hspace{0.01in}\text{d}X^{0}X^{A},\quad
X^{0}\hspace{0.01in}\text{d}X^{0}=b\,\text{d}X^{0}X^{0}. \label{BraidX0Dif}%
\end{equation}
We can see this if we apply the exterior derivative d to the relations in
Eq.~(\ref{BraidX0Dif}) and take into account the nilpotency and the Leibniz
rule for d:%
\begin{equation}
\text{d}^{2}=0,\qquad\text{d}(\rho\cdot\varsigma)=\text{d}\rho\cdot
\varsigma+(-1)^{n}\rho\cdot\text{d}\varsigma. \label{LeiRegd}%
\end{equation}
Note that the parameter $b$ in Eq.~(\ref{BraidX0Dif}) is undetermined and may
be any appropriate function of $q$.

As we know from the previous chapter, the R-ma\-trix $\hat{R}{^{\hspace
{0.01in}AB}}_{CD}$ for $\mathbb{R}_{q}^{3}$ is of block-dia\-go\-nal form [cf.
Eqs.~(\ref{RMatSOq3Anf})-(\ref{RMatSOq3End}) of the previous chapter].
Moreover, we know from Ref.~\cite{Lorek:1993tq} that adding the time element
$X^{0}$ to the Euclidean quantum space $\mathbb{R}_{q}^{3}$ extends the R-ma\-trix of
$\mathbb{R}_{q}^{3}$ by the following non-vanishing block:%
\begin{equation}%
\begin{tabular}
[c]{llllllll}
& $00$ & $0+$ & $03$ & $0-$ & $+0$ & $30$ & $-0$\\\cline{2-8}%
$00$ & \multicolumn{1}{|l}{$a$} & $0$ & $0$ & $0$ & $0$ & $0$ & $0$\\
$0+$ & \multicolumn{1}{|l}{$0$} & $0$ & $0$ & $0$ & $d$ & $0$ & $0$\\
$03$ & \multicolumn{1}{|l}{$0$} & $0$ & $0$ & $0$ & $0$ & $d$ & $0$\\
$0-$ & \multicolumn{1}{|l}{$0$} & $0$ & $0$ & $0$ & $0$ & $0$ & $d$\\
$+0$ & \multicolumn{1}{|l}{$0$} & $e$ & $0$ & $0$ & $0$ & $0$ & $0$\\
$30$ & \multicolumn{1}{|l}{$0$} & $0$ & $e$ & $0$ & $0$ & $0$ & $0$\\
$-0$ & \multicolumn{1}{|l}{$0$} & $0$ & $0$ & $e$ & $0$ & $0$ & $0$%
\end{tabular}
\ \ \ .
\end{equation}
The parameters $a$, $d$, and$\ e$ have been undetermined up to now. The
extended R-matrix again has the eigenvalues $1$, $q^{-6}$, and $-q^{-4}$. Its
additional block, however, leads to further eigenvalues $a$ and $\pm
\sqrt{d\hspace{0.01in}e}$. For each of these eigenvalues exists a projector to
the corresponding eigenspace. As before, the eigenvalues $1$, $q^{-6}$, and
$-q^{-4}$ refer to the projectors $P_{S}$, $P_{T}$, and $P_{A}$, which we have
already described in the previous chapter.\footnote{The new eigenvalues $a$
and $\pm\sqrt{d\hspace{0.01in}e}$ do not modify the non-vanishing entries of
the matrices representing the projectors $P_{S}$, $P_{T}$, and $P_{A}$.} We
denote the new projectors for the eigenvalues $\sqrt{d\hspace{0.01in}e}$ and
$-\sqrt{d\hspace{0.01in}e}$ by $P_{+}$ and $P_{-}$, respectively. To calculate
these projectors, we can use the following polynomials of the extended matrix
$\hat{R}$:%
\begin{align}
P_{+}  &  =\frac{(\hat{R}\hspace{0.01in}+\sqrt{d\hspace{0.01in}e}%
\operatorname*{id})(\hat{R}-\operatorname*{id})(\hat{R}\hspace{0.01in}%
-q^{-6}\operatorname*{id})(\hat{R}\hspace{0.01in}+q^{-4}\operatorname*{id}%
)(\hat{R}\hspace{0.01in}-a\operatorname*{id})}{2\sqrt{d\hspace{0.01in}%
e}\hspace{0.01in}(\sqrt{d\hspace{0.01in}e}-1)(\sqrt{d\hspace{0.01in}e}%
-q^{-6})(\sqrt{d\hspace{0.01in}e}+q^{-4})(\sqrt{d\hspace{0.01in}e}%
-a)},\\[0.1in]
P_{-}  &  =\frac{(\hat{R}\hspace{0.01in}-\sqrt{d\hspace{0.01in}e}%
\operatorname*{id})(\hat{R}-\operatorname*{id})(\hat{R}\hspace{0.01in}%
-q^{-6}\operatorname*{id})(\hat{R}\hspace{0.01in}+q^{-4}\operatorname*{id}%
)(\hat{R}\hspace{0.01in}-a\operatorname*{id})}{-2\sqrt{d\hspace{0.01in}%
e}\hspace{0.01in}(\sqrt{d\hspace{0.01in}e}+1)(\sqrt{d\hspace{0.01in}e}%
+q^{-6})(\sqrt{d\hspace{0.01in}e}-q^{-4})(\sqrt{d\hspace{0.01in}e}+a)}.
\end{align}
For the sake of completeness, we note that the matrix of the projector
referring to the eigenvalue $a$ takes on the following form:\footnote{We use
the convention that uppercase letters denote indices of spatial coordinates.
Lowercase letters denote indices of space-time coordinates.}%
\[
(P_{a}){^{ij}}_{kl}=\delta_{i0}\hspace{0.01in}\delta_{j0}\hspace{0.01in}%
\delta_{k0}\hspace{0.01in}\delta_{l0}.
\]

Remember that the projector $P_{A}$ defines the commutation relations for the
coordinates of $\mathcal{\mathbb{R}}_{q}^{3}$ [cf. Eq.~(\ref{ProASymKooRel})
of the previous chapter]. In addition to this, the new projector $P_{-}$ leads
us to the commutation relations between the time element $X^{0}$ and the three
coordinate generators $X^{+}$, $X^{3}$,$\ $and $X^{-}$ if we take into account
the condition $d=e$. In other words, the identities $(P_{-}){^{ij}}%
_{kl}\,X^{k}X^{l}\hspace{-0.01in}=0$ give the commutation relations of
Eq.~(\ref{VerRelZeiRauKooDreQua}) by setting $d=e$.

As we know, the projectors $P_{S}$ and $P_{T}$ determine the commutation
relations for the coordinate differentials d$X^{+}$, d$X^{3}$,$\ $and d$X^{-}%
$[cf. Eq.~(\ref{RelKooDif}) of the previous chapter]. Setting $d=e$, the new
projector $P_{+}$ determines the commutation relations between d$X^{0}$ and
the three coordinate differentials d$X^{+}$, d$X^{3}$,$\ $and d$X^{-}$, i.~e.
the identities $(P_{+}){^{ij}}_{kl}\,$d$X^{k}$d$X^{l}\hspace{-0.01in}=0$ give
the commutation relations in Eq.~(\ref{VerDiffXAX0ExtEuk}).

\subsection{Differential calculus\label{KapDifCal}}

To begin with, we calculate commutation relations between space-time
coordinates and their corresponding partial derivatives. Due to the Leibniz
rule of the external derivative, the following applies:%
\begin{equation}
\text{d}X^{j}.=(\text{d}X^{j})\hspace{0.01in}.+X^{j}\text{d}\hspace{0.01in}.
\label{ComRelExtDerCoo}%
\end{equation}
Note that the dots in the above identity indicate an unspecified element. We
can write the exterior derivative in terms of partial derivatives and
coordinate differentials:%
\begin{equation}
\text{d}=\text{d}X^{i}\partial_{i}=\text{d}X^{0}\partial_{0}+\text{d}%
X^{A}\partial_{A}.
\end{equation}
Plugging this into Eq.~(\ref{ComRelExtDerCoo}), we obtain the following
identities:%
\begin{align}
\text{d}X^{i}\hspace{0.01in}\partial_{i}X^{j}  &  =\text{d}X^{j}+X^{j}%
\hspace{0.01in}\text{d}X^{i}\partial_{i}\nonumber\\
&  =\text{d}X^{j}+X^{j}\hspace{0.01in}\text{d}X^{0}\partial_{0}+X^{j}%
\hspace{0.01in}\text{d}X^{A}\partial_{A}. \label{HerVerRelAblKoo}%
\end{align}
Using Eq.~(\ref{RelKooDifKoo}) of Chap.~\ref{KapQuaZeiEle} and
Eq.~(\ref{BraidX0Dif}) of the previous chapter, we can take all coordinate
differentials in the summands of Eq.~(\ref{HerVerRelAblKoo}) to the left:%
\begin{align}
\text{d}X^{i}\hspace{0.01in}\partial_{i}X^{A}  &  =\text{d}X^{0}\partial
_{0}X^{A}+\text{d}X^{B}\partial_{B}X^{A}=\text{d}X^{A}+X^{A}\hspace
{0.01in}\text{d}X^{0}\partial_{0}+X^{A}\hspace{0.01in}\text{d}X^{C}%
\partial_{C}\nonumber\\
&  =\text{d}X^{A}+\text{d}X^{0}X^{A}\hspace{0.01in}\partial_{0}+q^{4}\hat
{R}{^{AC}}_{BD}\,\text{d}X^{B}X^{D}\partial_{C},\\[0.05in]
\text{d}X^{i}\hspace{0.01in}\partial_{i}X^{0}  &  =\text{d}X^{0}\partial
_{0}X^{0}+\text{d}X^{A}\partial_{A}X^{0}=\text{d}X^{0}+X^{0}\hspace
{0.01in}\text{d}X^{0}\partial_{0}+X^{0}\hspace{0.01in}\text{d}X^{A}%
\partial_{A}\nonumber\\
&  =\text{d}X^{0}+b\hspace{0.01in}\text{d}X^{0}X^{0}\partial_{0}+\text{d}%
X^{A}X^{0}\partial_{A}.
\end{align}
If we compare the first expression with the last one in each of the two
calculations above and take into account the linear independence of the
coordinate differentials, we can read off the following Leibniz rules:%
\begin{align}
\partial_{B}X^{A}  &  =\delta_{B}^{A}+q^{4}\hat{R}{^{AC}}_{BD}\,X^{D}%
\partial_{C},\nonumber\\
\partial_{A}X^{0}  &  =X^{0}\hspace{0.01in}\partial_{A},\nonumber\\
\partial_{0}\hspace{0.01in}X^{A}  &  =X^{A}\hspace{0.01in}\partial
_{0},\nonumber\\
\partial_{0}\hspace{0.01in}X^{0}  &  =1+b\hspace{0.01in}X^{0}\hspace
{0.01in}\partial_{0}. \label{DifKalExtEukQuaDreUnk}%
\end{align}
The constant $b$ remains undetermined so that we choose $b=1$ for reasons of simplicity.

The partial derivatives of a $q$-deformed quantum space establish a quantum
space, again. This quantum space has the same algebraic structure as that of
the $q$-deformed space-time coordinates
\cite{CarowWatamura:1990zp,Wess:1990vh}. Thus, the $q$-deformed partial
derivatives $\partial_{i}$ commute with each other in the same way as the
covariant coordinate generators $X_{i}$:%
\begin{gather}
\partial_{0}\hspace{0.01in}\partial_{+}=\hspace{0.01in}\partial_{+}%
\hspace{0.01in}\partial_{0},\quad\partial_{0}\hspace{0.01in}\partial
_{-}=\hspace{0.01in}\partial_{-}\hspace{0.01in}\partial_{0},\quad\partial
_{0}\hspace{0.01in}\partial_{3}=\partial_{3}\hspace{0.01in}\partial
_{0},\nonumber\\
\partial_{+}\hspace{0.01in}\partial_{3}=q^{2}\partial_{3}\hspace
{0.01in}\partial_{+},\quad\partial_{3}\hspace{0.01in}\partial_{-}%
=\hspace{0.01in}q^{2}\partial_{-}\hspace{0.01in}\partial_{3},\nonumber\\
\partial_{+}\hspace{0.01in}\partial_{-}-\partial_{-}\hspace{0.01in}%
\partial_{+}=\hspace{0.01in}\lambda\hspace{0.01in}\partial_{3}\hspace
{0.01in}\partial_{3}.
\end{gather}
These commutation relations are invariant under conjugation if the derivatives
show the following conjugation properties:\footnote{The indices of the partial
derivatives are raised and lowered in the same way as those of the coordinates
[see Eq.~(\ref{HebSenInd}) in Chap.~\ref{KapQuaZeiEle}].}%
\begin{equation}
\overline{\partial_{A}}=-\hspace{0.01in}\partial^{A}=-g^{AB}\partial_{B},\qquad
\overline{\partial_{0}}=-\hspace{0.01in}\partial^{0}=-\hspace{0.01in}\partial_{0}. \label{KonAbl}%
\end{equation}
Conjugating the identities in Eq.~(\ref{DifKalExtEukQuaDreUnk}) yields the
Leibniz rules for another differential calculus. With $\hat{\partial}%
_{A}=q^{6}\partial_{A}$ and $\hat{\partial}_{0}=\partial_{0}$, we can write
the Leibniz rules of this second differential calculus in the following form:%
\begin{align}
\hat{\partial}_{B}\hspace{0.01in}X^{A}  &  =\delta_{B}^{A}+q^{-4}(\hat{R}%
^{-1}){^{AC}}_{BD}\,X^{D}\hat{\partial}_{C},\nonumber\\
\hat{\partial}_{A}\hspace{0.01in}X^{0}  &  =X^{0}\hspace{0.01in}\hat{\partial
}_{A},\nonumber\\
\hat{\partial}_{0}\hspace{0.01in}X^{A}  &  =X^{A}\hspace{0.01in}\hat{\partial
}_{0},\nonumber\\
\hat{\partial}_{0}\hspace{0.01in}X^{0}  &  =1+X^{0}\hspace{0.01in}%
\hat{\partial}_{0}. \label{DifKalExtEukQuaDreKon}%
\end{align}

\subsection{Hopf structures}

The algebra of the $q$-deformed partial derivatives $\partial^{A}$,
$A\in\{+,3,-\}$, together with $\mathcal{U}_{q}(\operatorname*{su}_{2})$ form
the cross-product algebra $\mathbb{R}_{q}^{3}\rtimes\hspace{0.01in}%
\mathcal{U}_{q}(\operatorname*{su}\nolimits_{2})$\footnote{From an algebraic
point of view, the $q$-deformed partial derivatives $\partial^{A}$ and the
coordinates $X^{A}$ behave in the same way.}. We know that the algebra
$\mathbb{R}_{q}^{3}\rtimes\mathcal{U}_{q}(\operatorname*{su}\nolimits_{2})$ is
a Hopf algebra \cite{Klimyk:1997eb}. Accordingly, the $q$-deformed partial
derivatives as elements of $\mathbb{R}_{q}^{3}\rtimes\hspace{0.01in}%
\mathcal{U}_{q}(\operatorname*{su}\nolimits_{2})$ have a co-product, an
antipode, and a co-unit. However, there are two ways of choosing the Hopf
structure of the $q$-deformed partial derivatives. It is so because the two
different co-products of the $q$-deformed partial derivatives are related to
the two versions of Leibniz rules given in Eq.~(\ref{DifKalExtEukQuaDreUnk})
or Eq.~(\ref{DifKalExtEukQuaDreKon}) of the last subchapter. To better
understand this, we note that you can generalize these Leibniz rules by
introducing so-called L-matrices $\mathcal{L}_{\partial}$ and $\mathcal{\bar
{L}}_{\partial}$ ($u\in\mathbb{R}_{q}^{3}$):%
\begin{align}
\partial^{A}u  &  =(\partial_{(1)}^{A}\triangleright u)\,\partial_{(2)}%
^{A}=\partial^{A}\triangleright u\hspace{0.01in}+\big ((\mathcal{L}_{\partial
}){^{A}}_{B}\triangleright u\big )\partial^{B},\nonumber\\
\hat{\partial}^{A}u  &  =(\hat{\partial}_{(\bar{1})}^{A}\,\bar{\triangleright
}\,u)\,\hat{\partial}_{(\bar{2})}^{A}=\hat{\partial}^{A}\,\bar{\triangleright
}\,u\hspace{0.01in}+\big ((\mathcal{\bar{L}}_{\partial}){^{A}}_{B}%
\triangleright u\big )\hat{\partial}^{B}. \label{AllVerRelParAblEle1}%
\end{align}
As you can see from the above identities, the two L-matrices determine the two
co-products of the $q$-deformed partial derivatives \cite{ogievetsky1992}:%
\begin{align}
\partial_{(1)}^{A}\otimes\partial_{(2)}^{A}  &  =\partial^{A}\otimes
1+(\mathcal{L}_{\partial}){^{A}}_{B}\otimes\partial^{B},\nonumber\\
\hat{\partial}_{(1)}^{A}\otimes\hat{\partial}_{(2)}^{A}  &  =\hat{\partial
}^{A}\otimes1+(\mathcal{\bar{L}}_{\partial}){^{A}}_{B}\otimes\hat{\partial
}^{B}. \label{KopParAllg}%
\end{align}
Note that the entries of the two L-matrices consist of generators of the Hopf
algebra $\mathcal{U}_{q}(\operatorname*{su}_{2})$ and powers of a scaling
operator $\Lambda$ [also see Eq.~(\ref{SkaOpeWir})]. For this reason, the
L-matrices can act on any element of $\mathbb{R}_{q}^{3}$.

In Ref.~\cite{Bauer:2003} and Ref.~\cite{Mikulovic:2006}, we have written down
the co-products of the partial derivatives $\partial^{A}$ or $\hat{\partial
}^{A}$, $A\in\{+,3,-\}$, explicitly. By taking into account
Eq.~(\ref{KopParAllg}), you can read off the entries of the L-matrices
$\mathcal{L}_{\partial}$ and $\mathcal{\bar{L}}_{\partial}$ from these
co-products. In doing so, you find, for example:%
\begin{equation}
(\mathcal{L}_{\partial}){^{-}}_{-}=\Lambda^{1/2}\hspace{0.01in}\tau
^{-1/2}\text{ und }(\mathcal{\bar{L}}_{\partial}){^{+}}_{+}=\Lambda
^{-1/2}\hspace{0.01in}\tau^{-1/2}.
\end{equation}
The scaling operator $\Lambda$ acts on the spatial coordinates or the
corresponding partial derivatives as follows:%
\begin{equation}
\Lambda\triangleright X^{A}=q^{4}X^{A},\qquad\Lambda\triangleright\partial
^{A}=q^{-4}\partial^{A}. \label{SkaOpeWir}%
\end{equation}
These actions imply the commutation relations%
\begin{equation}
\Lambda\hspace{0.01in}X^{A}=q^{4}X^{A}\Lambda,\qquad\Lambda\hspace
{0.01in}\partial^{A}=q^{-4}\partial^{A}\Lambda\label{VerSkaKooQuaEukDrei}%
\end{equation}
if we take into account the Hopf structure of $\Lambda$ \cite{ogievetsky1992}:%
\begin{equation}
\Delta(\Lambda)=\Lambda\otimes\Lambda,\qquad S(\Lambda)=\Lambda^{-1}%
,\qquad\varepsilon(\Lambda)=1. \label{HopStrLam}%
\end{equation}

Now, we modify the above considerations in such a way that they also apply to
the time element $X^{0}$ and the partial derivative $\partial^{0}$ or
$\hat{\partial}^{0}$. For this purpose, we extend the L-matrix $\mathcal{L}%
_{\partial}$ or $\mathcal{\bar{L}}_{\partial}$ so that we can also use it to
express the Leibniz rules for $\partial^{0}$ or $\hat{\partial}^{0}$ [see
Eq.~(\ref{DifKalExtEukQuaDreUnk}) or Eq.~(\ref{DifKalExtEukQuaDreKon}) of the
previous subchapter]:%
\begin{align}
\partial^{0}X^{i}  &  =(\partial_{(1)}^{0}\triangleright X^{i})\,\partial
_{(2)}^{0}=\partial^{0}\triangleright X^{i}\hspace{0.01in}+\big ((\mathcal{L}%
_{\partial}){^{0}}_{j}\triangleright X^{i}\big )\partial^{j}\nonumber\\
&  =\delta_{0}^{i}+X^{0}\partial_{0},\\[0.06in]
\hat{\partial}^{0}X^{i}  &  =(\hat{\partial}_{(\bar{1})}^{0}\,\bar
{\triangleright}\,X^{i})\,\hat{\partial}_{(\bar{2})}^{0}=\partial^{0}%
\,\bar{\triangleright}\,X^{i}\hspace{0.01in}+\big ((\mathcal{\bar{L}%
}_{\partial}){^{0}}_{j}\triangleright X^{i}\big )\hat{\partial}^{j}\nonumber\\
&  =\delta_{0}^{i}+X^{0}\hat{\partial}_{0}.
\end{align}
The identities above imply%
\begin{equation}
(\mathcal{L}_{\partial}){^{0}}_{j}=(\mathcal{\bar{L}}_{\partial}){^{0}}%
_{j}=\delta_{j}^{\hspace{0.01in}0}.
\end{equation}
Consequently, we have%
\begin{align}
\Delta(\partial^{0})  &  =\partial^{0}\otimes1+1\otimes\partial^{0}%
,\nonumber\\
\bar{\Delta}(\hat{\partial}^{0})  &  =\hat{\partial}^{0}\otimes1+1\otimes
\hat{\partial}^{0}. \label{CoProZeiAbl}%
\end{align}

Next, we look at the commutation relations between $\partial^{A}$ and $X^{0}%
$:\footnote{The following considerations also apply to the derivatives
$\hat{\partial}^{A}$ if we replace $\mathcal{L}_{\partial}$ by $\mathcal{\bar
{L}}_{\partial}$.}%
\begin{equation}
\partial^{A}X^{0}=\partial^{A}\triangleright X^{0}+\big ((\mathcal{L}%
_{\partial}){^{A}}_{B}\triangleright X^{0}\big )\partial^{B}=X^{0}\partial
^{A}.
\end{equation}
Thus, we get:%
\begin{equation}
(\mathcal{L}_{\partial}){^{A}}_{B}\triangleright X^{0}=\delta_{B}^{A}\,X^{0}.
\end{equation}
Since the entries of the L-matrices depend on the scaling operator $\Lambda$,
the above result requires that $\Lambda$ acts on $X^{0}$ as follows:%
\begin{equation}
\Lambda\triangleright X^{0}=X^{0}. \label{WirSkaZei}%
\end{equation}
If we take into account the Hopf structure of $\Lambda$ [cf.
Eq.~(\ref{HopStrLam})], Eq.~(\ref{WirSkaZei}) results in the following
commutation relation:%
\begin{equation}
\Lambda\hspace{0.01in}X^{0}=X^{0}\Lambda.
\end{equation}
To ensure that the existence of $\Lambda$ does not change the relation
$\partial_{0}\hspace{0.01in}X^{0}=1+X^{0}\partial_{0}$, we require that
$\Lambda$ commutates with $\partial^{0}$:%
\begin{equation}
\Lambda\hspace{0.01in}\partial^{0}=\partial^{0}\Lambda.
\end{equation}

The Hopf structure of the partial derivatives includes not only a co-product
but also an antipode and a co-unit. For the co-unit of the partial derivatives
applies \cite{ogievetsky1992}:%
\[
\varepsilon(\partial^{i})=0.
\]
We can obtain the antipodes of the partial derivatives from their co-products
using the following Hopf algebra axioms:%
\begin{equation}
a_{(1)}\cdot S(a_{(2)})=\varepsilon(a)=S(a_{(1)})\cdot a_{(2)}.
\end{equation}
Due to this axiom, we have:%
\begin{equation}
S(\partial^{i})=-\hspace{0.01in}\partial^{j}S^{-1}(\mathcal{L}_{\partial
}){^{i}}_{j},\qquad\bar{S}(\hat{\partial}^{i})=-\hspace{0.01in}\hat{\partial
}^{j}S^{-1}(\mathcal{\bar{L}}_{\partial}){^{i}}_{j}. \label{AlgForInvAntAbl}%
\end{equation}
This way, for example, we get the following expressions for the antipodes of
the partial derivatives $\partial^{-}$ and $\hat{\partial}^{+}$ (also see
Ref.~\cite{Bauer:2003}):%
\begin{equation}
S(\partial^{-})=-\Lambda^{-1/2}\hspace{0.01in}\tau^{1/2}\hspace{0.01in}%
\partial^{-},\qquad\bar{S}(\hat{\partial}^{+})=-\Lambda^{1/2}\hspace
{0.01in}\tau^{1/2}\hspace{0.01in}\hat{\partial}^{+}.
\end{equation}
For the antipodes of the time derivative, we obtain analogously:%
\begin{equation}
S(\partial^{0})=-\hspace{0.01in}\partial^{0},\qquad\bar{S}(\hat{\partial}^{0})=-\hspace{0.01in}\hat
{\partial}^{0}.
\end{equation}

We know, partial derivatives and coordinates behave in the same way from an
algebraic point of view. Accordingly, we can also specify co-products,
antipodes, and co-units for the time element:%
\begin{gather}
\Delta(X^{0})=\bar{\Delta}(X^{0})=X^{0}\otimes1+1\otimes X^{0},\nonumber\\
S(X^{0})=\bar{S}(X^{0})=-X^{0},\qquad\varepsilon(X^{0})=\bar{\varepsilon
}(X^{0})=0.
\end{gather}

\section{$q$-Analysis with time element\label{KapQAnaTim}}

\subsection{Star-products\label{KapStePro}}

We start with some general considerations on star-products. An $N$%
-di\-men\-sional quantum space is an algebra $V_{q}$ which is generated by
non-com\-mu\-ta\-tive coordinates $X^{i}$ with $i=1,\ldots,N$, i.~e. the
coordinates of the quantum space satisfy certain non-tri\-vial commutation
relations. The commutation relations of the quantum space coordinates generate
a two-si\-ded ideal $\mathcal{I}$, which is invariant under actions of the
Hopf algebra describing the symmetry of $V_{q}$. From this point of view, a
quantum space is a quotient algebra which is formed by the free algebra
$\mathbb{C}[X^{1},X^{2},\ldots,X^{N}]$ and the ideal $\mathcal{I}$:%
\begin{equation}
V_{q}=\frac{\mathbb{C}[X^{1},X^{2},\ldots,X^{N}]}{\mathcal{I}}.
\end{equation}

We can only prove the validity of a physical theory if it predicts measurement
results. The problem, however, is: How can we associate the elements of a
quantum space with real numbers? One solution to this problem is to interpret
the quantum space coordinates $X^{i}$ as operators acting on a ground state
which is invariant under actions of the symmetry Hopf algebra. This way, the
corresponding expectation values denoted as%
\begin{equation}
x^{i}=\langle\hspace{0.01in}X^{i}\rangle
\end{equation}
serve as real-valued variables with their numbers depending on the underlying
ground state. In the following, we will show how to extend the above identity
to normal-ordered monomials of quantum space coordinates.

The normal-ordered monomials of the quantum space coordinates $X^{i}$ form a
basis of the quantum space $V_{q}$, i.~e. we can write each element $F\in
V_{q}$ uniquely as a finite or infinite linear combination of monomials of a
given normal ordering (\textit{Poincar\'{e}-Birkhoff-Witt property}):%
\begin{equation}
F=\sum\limits_{i_{1},\ldots,\hspace{0.01in}i_{N}}a_{\hspace{0.01in}i_{1}%
\ldots\hspace{0.01in}i_{N}}\,(X^{1})^{i_{1}}\ldots\hspace{0.01in}%
(X^{N})^{i_{N}}\quad\text{with}\quad a_{\hspace{0.01in}i_{1}\ldots
\hspace{0.01in}i_{N}}\in\mathbb{C}.
\end{equation}
Since the monomials $(x^{1})^{i_{1}}\ldots\hspace{0.01in}(x^{N})^{i_{N}}$ with
$i_{1},\ldots,i_{N}\in\mathbb{N}_{0}$ form a basis of the commutative algebra
$V\mathcal{=\,}\mathbb{C}[\hspace{0.01in}x^{1},\ldots,x^{N}]$, we can define a
vector space isomorphism between $V$ and $V_{q}$, i.~e.%
\begin{equation}
\mathcal{W}:V\rightarrow V_{q} \label{VecRauIsoInv}%
\end{equation}
with%
\begin{equation}
\mathcal{W}\big ((x^{1})^{i_{1}}\ldots\hspace{0.01in}(x^{N})^{i_{N}%
}\big )=(X^{1})^{i_{1}}\ldots\hspace{0.01in}(X^{N})^{i_{N}}. \label{StePro0}%
\end{equation}
By linear extension follows%
\begin{equation}
V\ni f\mapsto F\in V_{q},
\end{equation}
where%
\begin{align}
f  &  =\sum\limits_{i_{1},\ldots,\hspace{0.01in}i_{N}}a_{\hspace{0.01in}%
i_{1}\ldots\hspace{0.01in}i_{N}}\,(x^{1})^{i_{1}}\ldots\hspace{0.01in}%
(x^{N})^{i_{N}},\nonumber\\
F  &  =\sum\limits_{i_{1},\ldots,\hspace{0.01in}i_{N}}a_{\hspace{0.01in}%
i_{1}\ldots\hspace{0.01in}i_{N}}\,(X^{1})^{i_{1}}\ldots\hspace{0.01in}%
(X^{N})^{i_{N}}. \label{AusFfNorOrd}%
\end{align}

The vector space isomorphism $\mathcal{W}$ is nothing else but the so-called
\textit{Mo\-yal-Weyl mapping}, which gives an operator $F\in V_{q}$ to a
complex-valued function $f\in V$ \cite{Bayen:1977ha, 1997q.alg.....9040K,
Madore:2000en, Moyal:1949sk}. You can see that the inverse of the Moyal-Weyl
mapping provides each quantum space coordinate with its expectation value:%
\begin{equation}
\mathcal{W}^{\hspace{0.01in}-1}(X^{i})=x^{i}=\langle X^{i}\rangle.
\end{equation}
This relation can be used for normal-ordered monomials as follows:%
\begin{align}
\mathcal{W}^{\hspace{0.01in}-1}((X^{1})^{i_{1}}\ldots\hspace{0.01in}%
(X^{N})^{i_{N}})  &  =(x^{1})^{i_{1}}\ldots\hspace{0.01in}(x^{N})^{i_{N}%
}=\langle(X^{1})^{i_{1}}\rangle\ldots\langle(X^{N})^{i_{N}}\rangle\nonumber\\
&  =\langle(X^{1})^{i_{1}}\ldots\hspace{0.01in}(X^{N})^{i_{N}}\rangle.
\end{align}
By linear extension, the vector space isomorphism $\mathcal{W}^{\hspace
{0.01in}-1}$ can assign an expectation value $f=\langle F\rangle$ to any
element $F$ of the quantum space $V_{q}$:%
\begin{equation}
\mathcal{W}^{\hspace{0.01in}-1}(F)=f=\langle F\rangle.
\end{equation}

Eq.~(\ref{AusFfNorOrd}) shows that the expectation value $f$ is a function of
the commutative coordinates $x^{i}$. This way, the vector space isomorphism
$\mathcal{W}^{\hspace{0.01in}-1}$ maps the non-commutative algebra $V_{q}$
onto the commutative algebra $V$ consisting of all power series with
coordinates $x^{i}$. We can even extend this vector space isomorphism to an
algebra isomorphism if we introduce a new product on the commutative algebra
$V$. This so-called \textit{star-product }symbolized by $\circledast$
satisfies the following homomorphism condition:%
\begin{equation}
\mathcal{W}^{\hspace{0.01in}-1}(F\cdot G)=\langle F\cdot G\rangle=\langle
F\rangle\circledast\langle G\hspace{0.01in}\rangle=\mathcal{W}^{\hspace
{0.01in}-1}(F)\circledast\mathcal{W}^{\hspace{0.01in}-1}(G). \label{HomMorBed}%
\end{equation}
With$\ f$ and $g$ as formal power series of the commutative coordinates
$x^{i}$, we can alternatively write the above condition in the following form:%
\begin{equation}
\mathcal{W}\left(  f\circledast g\right)  =\mathcal{W}\left(  f\right)
\cdot\mathcal{W}\left(  \hspace{0.01in}g\right)  . \label{HomBedWeyAbb}%
\end{equation}
Since the Mo\-yal-Weyl mapping is invertible, we can write the star-product as
follows:%
\begin{equation}
f\circledast g=\mathcal{W}^{\hspace{0.01in}-1}\big (\,\mathcal{W}\left(
f\right)  \cdot\mathcal{W}\left(  \hspace{0.01in}g\right)  \big ).
\label{ForStePro}%
\end{equation}
Thus, the star-product realizes the non-com\-mu\-ta\-tive product of $V_{q}$
on the commutative algebra $V$.

To get explicit formulas for calculating the star-product, we must define a
normal ordering for the non-commutative coordinate monomials. To derive these
formulas, we have to expand the non-com\-mu\-ta\-tive product of two
normal-ordered monomials in terms of normal-ordered monomials by using the
commutation relations for the quantum space coordinates:%
\begin{equation}
(X^{1})^{i_{1}}\ldots\hspace{0.01in}(X^{N})^{i_{N}}\cdot(X^{1})^{j_{1}}%
\ldots\hspace{0.01in}(X^{N})^{j_{N}}=\sum\nolimits_{k}B_{k}^{\hspace
{0.01in}i,j}\,(X^{1})^{k_{1}}\ldots\hspace{0.01in}(X^{N})^{k_{N}}.
\end{equation}

In the case of the Euclidean quantum space $\mathbb{R}_{q}^{3}$, it is useful
to determine the Moyal-Weyl mapping by the following choice of the normal-ordered
monomials ($n_{+},n_{3},n_{-}\in\mathbb{N}_{0}$):%
\begin{equation}
\mathcal{W}\big ((x^{+})^{n_{+}}(x^{3})^{n_{3}}(x^{-})^{n_{-}}\big )=(X^{+}%
)^{n_{+}}(X^{3})^{n_{3}}(X^{-})^{n_{-}}. \label{StePro1}%
\end{equation}
By using these normal-ordered monomials, we can obtain the following formula
to calculate the star-product ($\lambda=q-q^{-1}$):\ \footnote{For the
details, see Ref.~\cite{Wachter:2002A}.}$^{\text{,}}$\footnote{The argument
denoted by $\mathbf{x}$ indicates a dependence on the three spatial
coordinates $x^{+}$, $x^{3}$, and $x^{-}$.}%
\begin{equation}
f(\mathbf{x})\circledast g(\mathbf{x})=\sum_{k\hspace{0.01in}=\hspace
{0.01in}0}^{\infty}\lambda^{k}\hspace{0.01in}\frac{(x^{3})^{2k}}{[[k]]_{q^{4}%
}!}\,q^{2(\hat{n}_{3}\hat{n}_{+}^{\prime}+\,\hat{n}_{-}\hat{n}_{3}^{\prime}%
)}D_{q^{4},\hspace{0.01in}x^{-}}^{k}f(\mathbf{x})\,D_{q^{4},\hspace
{0.01in}x^{\prime+}}^{k}g(\mathbf{x}^{\prime})\big|_{x^{\prime}\rightarrow
\hspace{0.01in}x}. \label{StaProForExp}%
\end{equation}
Note that the above expression depends on the operators%
\begin{equation}
\hat{n}_{A}=x^{A}\frac{\partial}{\partial x^{A}}\quad\text{with}\quad
q^{\hat{n}_{A}}(x^{A})^{k}=q^{k}(x^{A})^{k}%
\end{equation}
and the so-called Jackson derivatives \cite{Jackson:1910yd}:%
\begin{equation}
D_{q^{k},\hspace{0.01in}x}\,f=\frac{f(q^{k}x)-f(x)}{q^{k}x-x}.
\end{equation}
Moreover, the $q$-factorials are defined by%
\[
\lbrack\lbrack n]]_{q}!=[[1]]_{q}\hspace{0.01in}[[2]]_{q}\ldots\lbrack\lbrack
n-1]]_{q}\hspace{0.01in}[[n]]_{q},\qquad\lbrack\lbrack0]]_{q}!=1.
\]

If we add a time element $X^{0}$ to the $q$-deformed Euclidean space
$\mathcal{\mathbb{R}}_{q}^{3}$, we can again specify a Moyal-Weyl mapping
between the extended quantum space algebra and the corresponding commutative
coordinate algebra.\footnote{In the following, $\mathcal{\mathbb{R}}_{q}%
^{3|t}$ denotes the Euclidean quantum space extended by a time element.} The
commutative coordinate algebra is now generated by the spatial coordinates
$x^{+}$, $x^{3}$, and $x^{-}$ as well as the time coordinate $t$. Accordingly,
we modify the Moyal-Weyl mapping as follows:%
\begin{equation}
\mathcal{W}\left(  (x^{+})^{n_{+}}(x^{3})^{n_{3}}(x^{-})^{n_{-}}%
t^{\hspace{0.01in}n_{0}}\right)  =(X^{+})^{n_{+}}(X^{3})^{n_{3}}(X^{-}%
)^{n_{-}}(X^{0})^{n_{0}}. \label{SteId1ExtEukQua}%
\end{equation}

Since the time element $X^{0}$ commutes with all coordinate generators of
$\mathcal{\mathbb{R}}_{q}^{3}$, $X^{0}$ does not modify the operator
expressions for the star-product, i.~e. the star-product formula in
Eq.~(\ref{StaProForExp}) still applies to commutative coordinate functions
which also depend on a time coordinate $t$. In other words, in
Eq.~(\ref{StaProForExp}) we can replace the two time-independent functions
$f(\mathbf{x})$ and $g(\mathbf{x})$ by the time-dependent functions
$f(\mathbf{x},t)$ and $g(\mathbf{x},t)$.

The algebra isomorphism $\mathcal{W}^{-1}$ can also be used to carry over the
conjugation of the quantum space algebra $V_{q}$ to the corresponding
commutative coordinate algebra $V$, i.~e. the mapping $\mathcal{W}%
^{\hspace{0.01in}-1}$ is a $\ast$-al\-ge\-bra homomorphism:%
\begin{equation}
\mathcal{W}(\hspace{0.01in}\overline{f}\hspace{0.01in})=\overline
{\mathcal{W}(f)}\qquad\Leftrightarrow\text{\qquad}\overline{f}=\mathcal{W}%
^{-1}\big (\hspace{0.01in}\overline{\mathcal{W}(f)}\hspace{0.01in}\big ).
\label{ConAlgIso}%
\end{equation}
The correspondence above implies the following conjugation property for the
star-pro\-duct:%
\begin{equation}
\overline{f\circledast g}=\overline{g}\circledast\overline{f}.
\label{KonEigSteProFkt}%
\end{equation}

Eq.~(\ref{ConSpaKoo})\ of Chap.~\ref{KapQuaZeiEle} and Eq.~(\ref{ConTim}) of
Chap.~\ref{KapComRel} imply that by conjugation a power series $f(\mathbf{x}%
,t)$ in the commutative space-time coordinates becomes (also see
Ref.~\cite{Wachter:2007A}):%
\begin{align}
\overline{f(\mathbf{x},t)}  &  =\sum\nolimits_{i}\bar{a}_{i_{+},i_{3}%
,i_{-},i_{0}}\,\overline{(\hspace{0.01in}x^{+})^{i_{+}}(\hspace{0.01in}%
x^{3})^{i_{3}}(\hspace{0.01in}x^{-})^{i_{-}}t^{i_{0}}}\nonumber\\
&  =\sum\nolimits_{i}\bar{a}_{i_{+},i_{3},i_{-},i_{0}}\,(-\hspace
{0.01in}q\hspace{0.01in}x^{-})^{i_{+}}(\hspace{0.01in}x^{3})^{i_{3}}%
(-\hspace{0.01in}q^{-1}x^{+})^{i_{-}}t^{i_{0}}\nonumber\\
&  =\sum\nolimits_{i}(-\hspace{0.01in}q)^{i_{+}-\hspace{0.01in}i_{-}}%
\hspace{0.01in}\bar{a}_{i_{-},i_{3},i_{+},i_{0}}\,(\hspace{0.01in}%
x^{+})^{i_{+}}(\hspace{0.01in}x^{3})^{i_{3}}(\hspace{0.01in}x^{-})^{i_{-}%
}t^{i_{0}}\nonumber\\
&  =\bar{f}(\mathbf{x},t). \label{KonPotReiKom}%
\end{align}
With $\bar{f}$, we designate the power series obtained from $f$ by
conjugation, and $\bar{a}_{i_{+},i_{3},i_{-},i_{0}}$ stands for the complex
conjugate of $a_{i_{+},i_{3},i_{-},i_{0}}$.

\subsection{Partial derivatives\label{ParAblKapAna}}

By using the Leibniz rules in Eq.$~$(\ref{DifKalExtEukQuaDreUnk}) or
Eq.$~$(\ref{DifKalExtEukQuaDreKon}) of Chap.~\ref{KapDifCal}, we can calculate
how the partial derivatives act on a normal-ordered monomial of
non-com\-mu\-ta\-tive coordinates. With the help of the Moyal-Weyl mapping,
these actions can be carried over to commutative coordinate monomials:%
\begin{equation}
\partial^{i}\triangleright(x^{+})^{n_{+}}(x^{3})^{n_{3}}(x^{-})^{n_{-}%
}t^{\hspace{0.01in}n_{0}}=\mathcal{W}^{\hspace{0.01in}-1}\big (\partial
^{i}\triangleright(X^{+})^{n_{+}}(X^{3})^{n_{3}}(X^{-})^{n_{-}}(X^{0})^{n_{0}%
}\big ).
\end{equation}
That the Mo\-yal-Weyl mapping is linear enables us to extend the action above
to space-time functions that can be expanded as a power series:%
\begin{equation}
\partial^{i}\triangleright f(\mathbf{x},t)=\mathcal{W}^{\hspace{0.01in}%
-1}\big (\partial^{i}\triangleright\mathcal{W}(f(\mathbf{x},t))\big ).
\end{equation}

If we use the ordering given in Eq.~(\ref{SteId1ExtEukQua}) of the previous
chapter, the Leibniz rules in Eq.~(\ref{DifKalExtEukQuaDreUnk}) of
Chap.~\ref{KapDifCal}\ lead to the following operator
representations:\footnote{$D_{q,\hspace{0.01in}x}^{n}$ denotes the $n$-fold
application of the Jackson derivative $D_{q,\hspace{0.01in}x}$.}%
\begin{align}
\partial_{+}\triangleright f(\mathbf{x},t)  &  =D_{q^{4},\hspace{0.01in}x^{+}%
}f(\mathbf{x},t),\nonumber\\
\partial_{3}\triangleright f(\mathbf{x},t)  &  =D_{q^{2},\hspace{0.01in}x^{3}%
}f(q^{2}x^{+},x^{3},x^{-},t),\nonumber\\
\partial_{-}\triangleright f(\mathbf{x},t)  &  =D_{q^{4},\hspace{0.01in}x^{-}%
}f(x^{+},q^{2}x^{3},x^{-},t)+\lambda\hspace{0.01in}x^{+}D_{q^{2}%
,\hspace{0.01in}x^{3}}^{2}f(\mathbf{x},t). \label{UnkOpeDarAbl}%
\end{align}
In Ref.~\cite{Bauer:2003}, we have already derived these representations for
time-\textit{in}dependent functions. That the representations above are also valid
for time-dependent functions results from the fact the time element $X^{0}$ commutes with the
spatial coordinates $X^{A}$ as well as the spatial derivatives $\partial_{A}$
[cf. Eq.~(\ref{DifKalExtEukQuaDreUnk}) of Chap.~\ref{KapDifCal}].

With the last two identities in Eq.~(\ref{DifKalExtEukQuaDreUnk}) of
Chap.~\ref{KapDifCal},\ we can determine how $\partial_{0}$ acts on
normal-ordered monomials of the generators $X^{i}$. Again, we can carry over
this action to the corresponding commutative coordinate algebra by the
isomorphism in Eq.~(\ref{SteId1ExtEukQua}) of the previous chapter. However, a
look at Eq.~(\ref{DifKalExtEukQuaDreUnk}) of Chap.~\ref{KapDifCal} shows that
$\partial_{0}$ commutes with the spatial coordinates $X^{A}$ and behaves like
an ordinary derivative with respect to $X^{0}$. Thus, $\partial_{0}$ is
realized on the commutative space-time algebra by an ordinary partial
derivative:%
\begin{equation}
\partial_{0}\triangleright\hspace{-0.01in}f(\mathbf{x},t)=\frac{\partial
f(\mathbf{x},t)}{\partial t}. \label{OpeDarZeiAblExtQuaEuk}%
\end{equation}

We can also use the Leibniz rules in Eq.$~$(\ref{DifKalExtEukQuaDreKon}) of
Chap.~\ref{KapDifCal} to derive operator representations for the partial
derivatives $\hat{\partial}_{i}$. To this end, we use normal-ordered monomials
different from those in Eq.~(\ref{SteId1ExtEukQua}) of the previous chapter:%
\begin{equation}
\widetilde{\mathcal{W}}\left(  t^{\hspace{0.01in}n_{0}}(x^{+})^{n_{+}}%
(x^{3})^{n_{3}}(x^{-})^{n_{-}}\right)  =(X^{0})^{n_{0}}(X^{-})^{n_{-}}%
(X^{3})^{n_{3}}(X^{+})^{n_{+}}.\label{UmNor}%
\end{equation}
If you compare the Leibniz rules in Eq.$~$(\ref{DifKalExtEukQuaDreUnk}) of
Chap.~\ref{KapDifCal} with those in Eq.$~$(\ref{DifKalExtEukQuaDreKon}) of the
same chapter, you can see that they transform into each other by the following
substitutions:%
\begin{gather}
q\rightarrow q^{-1},\quad X^{-}\rightarrow X^{+},\quad X^{+}\rightarrow
X^{-},\nonumber\\
\partial^{\hspace{0.01in}+}\rightarrow\hat{\partial}^{\hspace{0.01in}-}%
,\quad\partial^{\hspace{0.01in}-}\rightarrow\hat{\partial}^{\hspace{0.01in}%
+},\quad\partial^{\hspace{0.01in}3}\rightarrow\hat{\partial}^{\hspace
{0.01in}3},\quad\partial^{\hspace{0.01in}0}\rightarrow\hat{\partial}%
^{\hspace{0.01in}0}.\label{UebRegGedUngAblDreQua}%
\end{gather}
For this reason, we obtain the operator representations of the partial
derivatives $\hat{\partial}_{A}$ from those of the partial derivatives
$\partial_{A}$ [cf. Eq.~(\ref{UnkOpeDarAbl})] if we replace $q$ by $q^{-1}$
and exchange the indices $+$ and $-$:\footnote{To distinguish the conjugate
actions of partial derivatives from the unconjugate ones, we use
$\bar{\triangleright}$ to represent the conjugate actions. This distinction
also reminds us that the two actions refer to different normal-ordered
monomials.}%
\begin{align}
\hat{\partial}_{-}\,\bar{\triangleright}\,f(\mathbf{x},t) &  =D_{q^{-4}%
,\hspace{0.01in}x^{-}}f(\mathbf{x},t),\nonumber\\
\hat{\partial}_{3}\,\bar{\triangleright}\,f(\mathbf{x},t) &  =D_{q^{-2}%
,\hspace{0.01in}x^{3}}f(q^{-2}x^{-},x^{3},x^{+},t),\nonumber\\
\hat{\partial}_{+}\,\bar{\triangleright}\,f(\mathbf{x},t) &  =D_{q^{-4}%
,\hspace{0.01in}x^{+}}f(x^{-},q^{-2}x^{3},x^{+},t)-\lambda\hspace{0.01in}%
x^{-}D_{q^{-2},\hspace{0.01in}x^{3}}^{2}f(\mathbf{x},t).\label{KonOpeDarAbl}%
\end{align}

A look at the two last identities in Eq.~(\ref{DifKalExtEukQuaDreKon}) of
Chap.~\ref{KapDifCal} shows that the derivative $\hat{\partial}_{0}$ behaves
exactly like the derivative $\partial_{0}$. Accordingly, $\hat{\partial}_{0}$
is represented on the commutative space-time algebra by an ordinary partial
derivative, independent of the choice for the normal-ordered monomials:%
\begin{equation}
\hat{\partial}_{0}\,\bar{\triangleright}\,f(\mathbf{x},t)=\frac{\partial
f(\mathbf{x},t)}{\partial t}. \label{OpeDarZeiAblExtQuaEukKon}%
\end{equation}

For the sake of completeness, we mention that conjugation transforms
left-actions of partial derivatives into right-actions and vice versa:\footnote{You can calculate right actions of partial
derivatives by commuting a partial derivative from the right side of a
normal-ordered coordinate monomial to its left side using the Leibniz rules \cite{Bauer:2003}.}%
\begin{align}
\overline{\partial^{i}\triangleright f}  &  =-\bar{f}\,\bar{\triangleleft
}\,\partial_{i}, & \overline{f\,\bar{\triangleleft}\,\partial^{i}}  &
=-\hspace{0.01in}\partial_{i}\triangleright\bar{f},\nonumber\\
\overline{\hat{\partial}^{i}\,\bar{\triangleright}\,f}  &  =-\bar
{f}\triangleleft\hat{\partial}_{i}, & \overline{f\triangleleft\hat{\partial
}^{i}}  &  =-\hspace{0.01in}\hat{\partial}_{i}\,\bar{\triangleright}\,\bar{f}.
\label{RegConAbl}%
\end{align}
This fact implies that the right-action of $\partial_{0}$ or $\hat{\partial
}_{0}$ differs from its left-action by a minus sign, only:%
\begin{equation}
f(\mathbf{x},t)\triangleleft\hat{\partial}_{0}=\,\,f(\mathbf{x},t)\,\bar
{\triangleleft}\,\partial_{0}=-\frac{\partial f(\mathbf{x},t)}{\partial t}.
\end{equation}

\subsection{Integration\label{KapIntegral}}

Eqs.~(\ref{UnkOpeDarAbl}) and (\ref{KonOpeDarAbl}) of the previous chapter
show that the operator representations of $q$-de\-formed partial derivatives
consist of a term $\partial_{\operatorname*{cla}}^{A}$ and a so-called
correction term $\partial_{\operatorname*{cor}}^{A}$:%
\begin{equation}
\partial^{A}\triangleright F=\left(  \partial_{\operatorname*{cla}}%
^{A}+\partial_{\operatorname*{cor}}^{A}\right)  \triangleright F.
\end{equation}
The term $\partial_{\operatorname*{cla}}^{A}$ becomes an ordinary partial
derivative in the undeformed limit $q\rightarrow1$ and the term $\partial
_{\operatorname*{cor}}^{A}$ disappears in the undeformed limit. We can get a
solution to the difference equation $\partial^{A}\triangleright F=f$ with
given $f$ by using the following formula:%
\begin{align}
F  &  =(\partial^{A})^{-1}\triangleright f=\left(  \partial
_{\operatorname*{cla}}^{A}+\partial_{\operatorname*{cor}}^{A}\right)
^{-1}\triangleright f\nonumber\\
&  =\sum_{k\hspace{0.01in}=\hspace{0.01in}0}^{\infty}\left[  -(\partial
_{\operatorname*{cla}}^{A})^{-1}\partial_{\operatorname*{cor}}^{A}\right]
^{k}(\partial_{\operatorname*{cla}}^{A})^{-1}\triangleright f.
\end{align}
If we apply the above formula to the operator representations from
Eq.~(\ref{UnkOpeDarAbl}) in the previous chapter, we obtain\footnote{The
calculation of the operator expressions for $(\partial_{A})^{-1}$ remains the
same as in Ref.~\cite{Wachter:2004A} since the time element is independent of
the space coordinates.}%
\begin{align}
(\partial_{+})^{-1}\triangleright f(\mathbf{x},t)  &  =D_{q^{4},\hspace
{0.01in}x^{+}}^{-1}f(\mathbf{x},t),\nonumber\\
(\partial_{3})^{-1}\triangleright f(\mathbf{x},t)  &  =D_{q^{2},\hspace
{0.01in}x^{3}}^{-1}f(q^{-2}x^{+},x^{3},x^{-},t),
\end{align}
and%
\begin{gather}
(\partial_{-})^{-1}\triangleright f(\mathbf{x},t)=\nonumber\\
=\sum_{k\hspace{0.01in}=\hspace{0.01in}0}^{\infty}q^{2k\left(  k\hspace
{0.01in}+1\right)  }\left(  -\lambda\,x^{+}D_{q^{4},\hspace{0.01in}x^{-}}%
^{-1}D_{q^{2},\hspace{0.01in}x^{3}}^{2}\right)  ^{k}D_{q^{4},\hspace
{0.01in}x^{-}}^{-1}f(x^{+},q^{-2\left(  k\hspace{0.01in}+1\right)  }%
x^{3},x^{-},t).
\end{gather}
Note that $D_{q,\hspace{0.01in}x}^{-1}$ stands for a Jackson integral with $x$
being the variable of integration \cite{Jackson:1908}. The explicit form of
this Jackson integral depends on its limits of integration and the value for
the deformation parameter $q$. If $x>0$ and $q>1$, for example, the following
applies:%
\begin{align}
\int_{0}^{\hspace{0.01in}x}\text{d}_{q}z\hspace{0.01in}f(z)  &  =(q-1)\hspace
{0.01in}x\sum_{j=1}^{\infty}q^{-j}f(q^{-j}x),\nonumber\\
\int_{x}^{\hspace{0.01in}\infty}\text{d}_{q}z\hspace{0.01in}f(z)  &
=(q-1)\hspace{0.01in}x\sum_{j=0}^{\infty}q^{\hspace{0.01in}j}f(q^{\hspace
{0.01in}j}x).
\end{align}

By successively applying the $q$-in\-te\-grals for the different coordinates,
we can explain an integration over the entire position space. Apart from a
normalization factor, this integration is independent of the order in which we
apply the $q$-in\-te\-grals for the different coordinates
\cite{Wachter:2004A, Wachter:2007A}:%
\begin{equation}
\int_{-\infty}^{+\infty}\text{d}_{q}^{3}x\,f(\mathbf{x},t)=(\partial_{-}%
)^{-1}\big |_{-\infty}^{+\infty}\,(\partial_{3})^{-1}\big |_{-\infty}%
^{+\infty}\,(\partial_{+})^{-1}\big |_{-\infty}^{+\infty}\triangleright
f(\mathbf{x},t). \label{qIntEukQua}%
\end{equation}
On the right-hand side of the above relation, we can reduce the $q$%
-in\-te\-grals for the different coordinates to Jackson integrals:\footnote{This simplification results from the fact that
the function to be integrated must disappear at infinity \cite{Wachter:2004A}.}%
\begin{equation}
\int_{-\infty}^{+\infty}\text{d}_{q}^{3}x\,f(\mathbf{x},t)=D_{q^{2}%
,\hspace{0.01in}x^{-}}^{-1}\big |_{-\infty}^{+\infty}\,D_{q,x^{3}}%
^{-1}\big |_{-\infty}^{+\infty}\,D_{q^{2},\hspace{0.01in}x^{+}}^{-1}%
\big |_{-\infty}^{+\infty}\,f(\mathbf{x},t).
\end{equation}

For the sake of completeness, we mention that the $q$-integral over the entire
$q$-deformed Euclidean space behaves under quantum space conjugation as
follows:%
\begin{equation}
\overline{\int_{-\infty}^{+\infty}\text{d}_{q}^{3}x\,f(\mathbf{x},t)}%
=\int_{-\infty}^{+\infty}\text{d}_{q}^{3}x\,\overline{f(\mathbf{x},t)}.
\label{KonEigVolInt}%
\end{equation}
We outline how to prove the above identity. From the conjugation properties of
the partial derivatives follows [cf. Eq.~(\ref{KonAbl}) in
Chap.~\ref{KapDifCal}]:%
\begin{equation}
1=\overline{(\partial_{A})^{-1}\partial_{A}}=(-g^{AB}\partial_{B}%
)\,\overline{(\partial_{A})^{-1}}\quad\Rightarrow\quad\overline{(\partial
_{A})^{-1}}=-\frac{1}{g^{AB}}(\partial_{B})^{-1}.
\end{equation}
With this result, we can conjugate the expression in Eq.~(\ref{qIntEukQua}).
Doing so, we take into account that the quantum space conjugation transforms
left-actions of the integral operators into right-actions:%
\begin{align}
&  \overline{(\partial_{-})^{-1}\big |_{-\infty}^{+\infty}\,(\partial
_{3})^{-1}\big |_{-\infty}^{+\infty}\,(\partial_{+})^{-1}\big |_{-\infty
}^{+\infty}\triangleright f(\mathbf{x},t)}=\nonumber\\
&  \qquad\qquad=\overline{f(\mathbf{x},t)}\,\bar{\triangleleft}\,\overline
{(\partial_{+})^{-1}}\big |_{-\infty}^{+\infty}\,\overline{(\partial_{3}%
)^{-1}}\big |_{-\infty}^{+\infty}\,\overline{(\partial_{-})^{-1}%
}\big |_{-\infty}^{+\infty}\nonumber\\
&  \qquad\qquad=-\overline{f(\mathbf{x},t)}\,\bar{\triangleleft}%
\,(\partial_{-})^{-1}\big |_{-\infty}^{+\infty}\,(\partial_{3})^{-1}%
\big |_{-\infty}^{+\infty}\,(\partial_{+})^{-1}\big |_{-\infty}^{+\infty}.
\label{KonLinInt}%
\end{align}
If we express the right-actions of the elements $(\partial_{A})^{-1}$ by
Jackson integrals, we can see:%
\begin{align}
&  f(\mathbf{x},t)\,\bar{\triangleleft}\,(\partial_{-})^{-1}\big |_{-\infty
}^{+\infty}\,(\partial_{3})^{-1}\big |_{-\infty}^{+\infty}\,(\partial
_{+})^{-1}\big |_{-\infty}^{+\infty}=\nonumber\\
&  \qquad\qquad=-D_{q^{2},\hspace{0.01in}x^{-}}^{-1}\big |_{-\infty}^{+\infty
}\,D_{q,x^{3}}^{-1}\big |_{-\infty}^{+\infty}\,D_{q^{2},\hspace{0.01in}x^{+}%
}^{-1}\big |_{-\infty}^{+\infty}\,f(\mathbf{x},t)\nonumber\\
&  \qquad\qquad=-\int_{-\infty}^{+\infty}\text{d}_{q}^{3}x\,f(\mathbf{x},t).
\label{KonRechDarInt}%
\end{align}
Since the two signs in Eq.~(\ref{KonLinInt}) and Eq.~(\ref{KonRechDarInt})
cancel each other out, the identity in Eq.~(\ref{KonEigVolInt}) is established.

Not only can we add the central element $\partial_{0}$ to the algebra of
$q$-deformed partial derivatives, but also its inverse $(\partial_{0})^{-1}$.
Remember that $\partial_{0}$ acts on the commutative space-time algebra like
an ordinary partial derivative [cf. Eq.~(\ref{OpeDarZeiAblExtQuaEuk}) of the
previous chapter]. For this reason, the action of $(\partial_{0})^{-1}$ on a
commutative space-time function is nothing else but an ordinary integral:%
\begin{equation}
(\partial_{0})^{-1}\triangleright f(\mathbf{x},t)\hspace{0.01in}=\int
\text{d}t\,f(\mathbf{x},t).
\end{equation}

The above considerations also apply to the representations of the partial
derivatives $\hat{\partial}_{i}$ [cf. Eq.~\ref{KonOpeDarAbl} of the previous
chapter]. We know, however, that these representations follow from those of
the derivatives $\partial_{i}$ if we replace $q$ by $q^{-1}$ and exchange the
indices $+$ and $-$. Hence, if we apply these substitutions to the results of
this chapter, we immediately obtain the corresponding results for the partial
derivatives $\hat{\partial}_{i}$.

\subsection{Translations\label{KapTra}}

We start with some general considerations about translations on $q$-de\-formed
quantum spaces. For translations on $q$-de\-formed quantum spaces, we replace
every coordinate generator $X^{i}$ of a $q$-de\-formed quantum space $V_{q}$
by $X^{i}\otimes1+1\otimes Y^{i}$, where $Y^{i}$ denotes the coordinate
generator of a second $q$-de\-formed quantum space $V_{q}$
\cite{Chryssomalakos:1993zm, majid-1993-34}. This way, we get a mapping from
$V_{q}$ to the tensor product $V_{q}\otimes V_{q}$.

Since we can write each element of $V_{q}$ in terms of normal-ordered
monomials, you must only know how normal-ordered monomials behave under
translations. If we apply the above substitutions to any normal-ordered
monomial of quantum space coordinates, we obtain expressions that we can write
in terms of tensor products of two normal-ordered monomials:%
\begin{gather}
(X^{1}\otimes1+1\otimes Y^{1})^{i_{1}}\ldots(X^{N}\otimes1+1\otimes
Y^{N})^{i_{N}}=\nonumber\\
=\sum\nolimits_{k,l}\alpha_{i;k,l}\,(X^{1})^{k_{1}}\ldots(X^{N})^{k_{N}%
}\otimes(Y^{1})^{l_{1}}\ldots(Y^{N})^{l_{N}}. \label{TraExpGen}%
\end{gather}
To get the expansion above, you need the braiding relations between coordinate
generators of different quantum spaces [see Eq.~(\ref{ZopRel}) of
Chap.~\ref{KapQuaZeiEle}] as well as the commutation relations for coordinate
generators of the same quantum space.

Since all non-com\-mu\-ta\-tive monomials are normal-ordered in the expansion
above, we can carry over the identity in Eq.~(\ref{TraExpGen})\ to commutative
coordinate monomials. This way, we get a $q$-analog of the multidimensional
binomial formula. Then we can directly read off an operator representation
from this $q$-deformed formula. This operator representation enables us to
calculate $q$-de\-formed translations for all those functions which we can
write as a power series in the commutative coordinates $x^{i}$. In the case of
the three-dimensional $q$-de\-formed Euclidean space, for example, we can get
the following formula for calculating $q$-trans\-la\-tions
\cite{Wachter:2004phengl}:%
\begin{align}
f(\mathbf{x}\oplus\mathbf{y})=  &  \sum_{i_{+}=\hspace{0.01in}0}^{\infty}%
\sum_{i_{3}=\hspace{0.01in}0}^{\infty}\sum_{i_{-}=\hspace{0.01in}0}^{\infty
}\sum_{k\hspace{0.01in}=\hspace{0.01in}0}^{i_{3}}\frac{(-q^{-1}\lambda
\lambda_{+})^{k}}{[[2k]]_{q^{-2}}!!}\frac{(x^{-})^{i_{-}}(x^{3})^{i_{3}%
-\hspace{0.01in}k}(x^{+})^{i_{+}+\hspace{0.01in}k}\,(y^{-})^{k}}%
{[[i_{-}]]_{q^{-4}}!\,[[i_{3}-k]]_{q^{-2}}!\,[[i_{+}]]_{q^{-4}}!}\nonumber\\
&  \qquad\times\big (D_{q^{-4},\hspace{0.01in}y^{-}}^{i_{-}}D_{q^{-2}%
,\hspace{0.01in}y^{3}}^{i_{3}+\hspace{0.01in}k}\hspace{0.01in}D_{q^{-4}%
,\hspace{0.01in}y^{+}}^{i_{+}}f\big )(q^{2(k\hspace{0.01in}-\hspace
{0.01in}i_{3})}y^{-},q^{-2i_{+}}y^{3}).
\end{align}

As mentioned above, we derive $q$-trans\-la\-tions with the help of the
braiding relations for generators of different quantum spaces. However, there
are two ways of choosing these braiding relations [see also Eqs.~(\ref{ZopRel}%
) and (\ref{InvZopRel}) of Chap.~\ref{KapQuaZeiEle}]. Accordingly, there are
two versions of $q$-trans\-la\-tions on each $q$-de\-formed quantum space.
Whenever we want, we can transform the operator representations of the two
$q$-translations into each other by simple substitutions (see
Ref.~\cite{Wachter:2007A}).

The $q$-de\-formed quantum spaces we have considered so far are so-called
braided Hopf algebras \cite{Majid:1996kd}. From this point of view, the two
versions of $q$-trans\-lations are nothing else but realizations of two
braided co-pro\-ducts $\underline{\Delta}$ and $\underline{\bar{\Delta}}$ on
the corresponding commutative coordinate algebras \cite{Wachter:2007A}:%
\begin{align}
f(x\oplus y)  &  =((\mathcal{W}^{\hspace{0.01in}-1}\otimes\mathcal{W}%
^{\hspace{0.01in}-1})\circ\underline{\Delta})(\mathcal{W}%
(f)),\nonumber\\[0.08in]
f(x\,\bar{\oplus}\,y)  &  =((\mathcal{W}^{\hspace{0.01in}-1}\otimes
\mathcal{W}^{-1})\circ\underline{\bar{\Delta}})(\mathcal{W}(f)).
\label{KonReaBraCop}%
\end{align}

The braided Hopf algebras under consideration also have braided antipodes
$\underline{S}$ and $\bar{S}$, which can be realized on the corresponding
commutative coordinate algebras as well:%
\begin{align}
f(\ominus\,x)  &  =(\mathcal{W}^{\hspace{0.01in}-1}\circ\underline{S}%
\hspace{0.01in})(\mathcal{W}(f)),\nonumber\\
f(\bar{\ominus}\,x)  &  =(\mathcal{W}^{\hspace{0.01in}-1}\circ\underline
{\bar{S}}\hspace{0.01in})(\mathcal{W}(f)). \label{qInvDef}%
\end{align}
In the following, we refer to the operations in Eq.~(\ref{qInvDef})\ as
$q$\textit{-in\-ver\-sions}. In the case of the $q$-de\-formed Euclidean
space, for example, we can find the following operator representation for
$q$-in\-ver\-sions \cite{Wachter:2004phengl}:%
\begin{align}
\hat{U}^{-1}f(\ominus\,\mathbf{x})=  &  \sum_{i=0}^{\infty}(-q\lambda
\lambda_{+})^{i}\,\frac{(x^{+}x^{-})^{i}}{[[2i]]_{q^{-2}}!!}\,q^{-2\hat{n}%
_{+}(\hat{n}_{+}+\hspace{0.01in}\hat{n}_{3})-2\hat{n}_{-}(\hat{n}_{-}%
+\hspace{0.01in}\hat{n}_{3})-\hat{n}_{3}\hat{n}_{3}}\nonumber\\
&  \qquad\times D_{q^{-2},\hspace{0.01in}x^{3}}^{2i}\,f(-q^{2-4i}%
x^{-},-q^{1-2i}x^{3},-q^{2-4i}x^{+}).
\end{align}
Note that the operators $\hat{U}$ and $\hat{U}^{-1}$ act on a commutative
function $f(x^{+},x^{3},x^{-})$ as follows:%
\begin{align}
\hat{U}f  &  =\sum_{k\hspace{0.01in}=\hspace{0.01in}0}^{\infty}\left(
-\lambda\right)  ^{k}\frac{(x^{3})^{2k}}{[[k]]_{q^{-4}}!}\,q^{-2\hat{n}%
_{3}(\hat{n}_{+}+\hspace{0.01in}\hat{n}_{-}+\hspace{0.01in}k)}D_{q^{-4}%
,\hspace{0.01in}x^{+}}^{k}D_{q^{-4},\hspace{0.01in}x^{-}}^{k}f,\nonumber\\
\hat{U}^{-1}f  &  =\sum_{k\hspace{0.01in}=\hspace{0.01in}0}^{\infty}%
\lambda^{k}\hspace{0.01in}\frac{(x^{3})^{2k}}{[[k]]_{q^{4}}!}\,q^{2\hat{n}%
_{3}(\hat{n}_{+}+\hspace{0.01in}\hat{n}_{-}+\hspace{0.01in}k)}D_{q^{4}%
,\hspace{0.01in}x^{+}}^{k}D_{q^{4},\hspace{0.01in}x^{-}}^{k}f.
\end{align}

Due to its trivial braiding properties, the time element is independent of the
spatial coordinates. For this reason, displacements in time are independent of
translations in space. Since the time element behaves like a commutative
parameter, we can write displacements in time as an ordinary Taylor expansion:%
\begin{align}
f(\mathbf{x}\oplus\mathbf{y},t\oplus t^{\hspace{0.01in}\prime})  &
=\sum_{k\hspace{0.01in}=\hspace{0.01in}0}^{\infty}\frac{(t^{\hspace
{0.01in}\prime})^{k}}{k!}\,\frac{\partial f(\mathbf{x}\oplus\mathbf{y}%
,t)}{\partial t}=f(\mathbf{x}\oplus\mathbf{y},t+t^{\hspace{0.01in}\prime
}),\nonumber\\
f(\mathbf{x}\,\bar{\oplus}\,\mathbf{y},t\,\bar{\oplus}\,t^{\hspace
{0.01in}\prime})  &  =\sum_{k\hspace{0.01in}=\hspace{0.01in}0}^{\infty}%
\frac{(t^{\hspace{0.01in}\prime})^{k}}{k!}\,\frac{\partial f(\mathbf{x}%
\,\bar{\oplus}\,\mathbf{y},t)}{\partial t}=f(\mathbf{x}\,\bar{\oplus
}\,\mathbf{y},t+t^{\hspace{0.01in}\prime}).
\end{align}
Consequently, inversions in time are nothing else but a substitution of the
time coordinate by the negative one:%
\begin{equation}
f(\ominus\,\mathbf{x},\ominus\,t)=f(\ominus\,\mathbf{x},-\hspace
{0.01in}t),\qquad f(\bar{\ominus}\,\mathbf{x},\bar{\ominus}\,t)=f(\bar
{\ominus}\,\mathbf{x},-\hspace{0.01in}t).
\end{equation}

\subsection{Exponentials\label{KapExp}}

\begin{figure}
\begin{center}
\includegraphics[width=0.42\textwidth]{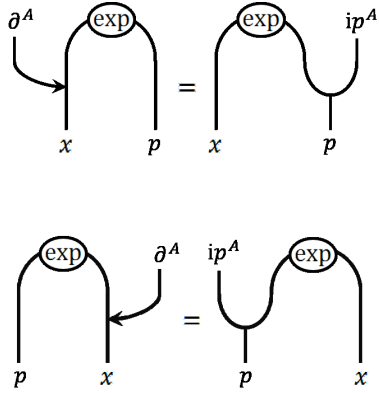}
\caption{Eigenvalue equations of $q$-exponentials}
\label{Fig1}       % Give a unique label
\end{center}
\end{figure}

A $q$-de\-formed exponential is an eigenfunction of each partial derivative of
a given $q$-de\-formed quantum space \cite{Majid:1993ud, Schirrmacher:1995,
Wachter:2004ExpA}. In the following, we consider $q$-de\-formed exponentials
that are eigenfunctions for left-actions or right-actions of partial
derivatives:%
\begin{align}
\text{i}^{-1}\partial^{A}\triangleright\exp_{q}(x|\text{i}p)  &  =\exp
_{q}(x|\text{i}p)\circledast p^{A},\nonumber\\
\exp_{q}(\text{i}^{-1}p|x)\,\bar{\triangleleft}\,\partial^{A}\text{i}^{-1}  &
=p^{A}\circledast\exp_{q}(\text{i}^{-1}p|x). \label{EigGl1N}%
\end{align}
For a better understanding, the above eigenvalue equations are shown
graphically in Fig.~\ref{Fig1}. The $q$-ex\-po\-nen\-tials are uniquely
defined by their eigenvalue equations in connection with the following
normalization conditions:%
\begin{align}
\exp_{q}(x|\text{i}p)|_{x\hspace{0.01in}=\hspace{0.01in}0}  &  =\exp
_{q}(x|\text{i}p)|_{p\hspace{0.01in}=\hspace{0.01in}0}=1,\nonumber\\
\exp_{q}(\text{i}^{-1}p|x)|_{x\hspace{0.01in}=\hspace{0.01in}0}  &  =\exp
_{q}(\text{i}^{-1}p|x)|_{p\hspace{0.01in}=\hspace{0.01in}0}=1.
\end{align}%
%TCIMACRO{\FRAME{ftbpFU}{4.5057in}{0.9141in}{0pt}{\Qcb{Eigenvalue equations of
%$q$-exponentials.}}{\Qlb{Fig1}}{fig1.jpg}%
%width 4.5057in;  height 0.9141in;  depth 0pt;  original-width 9.7464in;
%original-height 1.9527in;  cropleft "0";  croptop "1";  cropright "1";
%filename '../HabilArtikel/Figures/Fig1.jpg';file-properties "XNPEU";}} }%
%BeginExpansion

%\begin{figure}
%[ptb]
%\begin{center}
%\includegraphics[
%natheight=1.952700in,
%natwidth=9.746400in,
%height=0.9141in,
%width=4.5057in
%]%
%{Fig1.eps}%
%\caption{Eigenvalue equations of $q$-exponentials.}%
%\label{Fig1}%
%\end{center}
%\end{figure}
%%EndExpansion

To get explicit formulas for $q$-ex\-ponentials, we best consider the dual
pairings between the coordinate algebra of the $q$-de\-formed position space
and that of the corresponding $q$-de\-formed momentum space
\cite{Majid:1993ud}:%
\begin{align}
\big\langle f(\hspace{0.01in}p),g(x)\big\rangle_{p,\bar{x}}  & =
\lbrack\hspace{0.01in}f(\text{i}^{-1}\partial)\triangleright g(x)]_{x\hspace
{0.01in}=\hspace{0.01in}0},\nonumber\\
\big\langle f(x),g(\hspace{0.01in}p)\big\rangle_{x,\bar{p}}  &  =
\lbrack\hspace{0.01in}f(x)\,\bar{\triangleleft}\,g(\partial\hspace
{0.01in}\text{i}^{-1})]_{x\hspace{0.01in}=\hspace{0.01in}0}.
\end{align}
Let $\{e^{a}\}$ be a basis of the $q$-de\-formed position space algebra and
let $\{f_{b}\}$ be a dual basis of the corresponding $q$-de\-formed momentum
algebra:\footnote{$\mathcal{W}_{x}$ and $\mathcal{W}_{p}$ denote the
Moyal-Weyl mapping for the $q$-de\-formed position space algebra and that for
the $q$-de\-formed momentum space algebra, respectively.}%
\begin{equation}
\big\langle\mathcal{W}_{p}(f_{b}),\mathcal{W}_{x}(e^{a})\big\rangle_{p,\bar
{x}}=\delta_{b}^{a},\qquad\big\langle\mathcal{W}_{x}(e^{a}),\mathcal{W}%
_{p}(f_{b})\big\rangle_{x,\bar{p}}=\delta_{b}^{a},
\end{equation}
Now, we are able to write the $q$-ex\-po\-nen\-tials as canonical
elements:%
\begin{align}
\exp_{q}(x|\text{i}p)  &  = \sum\nolimits_{a}\mathcal{W}_{x}(e^{a}%
)\otimes\mathcal{W}_{p}(f_{a}),\nonumber\\
\exp_{q}(\text{i}p|x)  & = \sum\nolimits_{a}\mathcal{W}_{p}(f_{a}%
)\otimes\mathcal{W}_{x}(e^{a}). \label{GenForQExp}%
\end{align}

The monomials $t^{\hspace{0.01in}m_{0}}(x^{+})^{m_{+}}(x^{3})^{m_{3}}(x^{-})^{m_{-}}$ form a
basis of the commutative coordinate algebra corresponding to the Euclidean quantum space with a time element. We can obtain the elements of the dual basis
by the action of the partial derivatives on these monomials. With the help of
the operator representations for the partial derivatives in
Eqs.~(\ref{UnkOpeDarAbl}) and (\ref{OpeDarZeiAblExtQuaEuk}) of
Chap.~\ref{ParAblKapAna}, you can find:\footnote{The corresponding expressions
for the derivatives $\hat{\partial}_{i}$ can be obtained by replacing $q$ with
$q^{-1}$ and exchanging the indices $+$ and $-$.}%
\begin{gather}
\big \langle(\text{i}^{-1}E)^{n_{0}}(\text{i}p_{-})^{n_{-}}(\text{i}%
p_{3})^{n_{3}}(\text{i}p_{+})^{n_{+}},t^{\hspace{0.01in}m_{0}}(x^{+})^{m_{+}}(x^{3})^{m_{3}%
}(x^{-})^{m_{-}}\big \rangle_{p,\hspace{0.01in}\bar{x}}=\nonumber\\
=\big [(\partial_{0})^{n_{0}}(\partial_{-})^{n_{-}}(\partial_{3})^{n_{3}%
}(\partial_{+})^{n_{+}}\triangleright t^{\hspace{0.01in}m_{0}}(x^{+})^{m_{+}}(x^{3})^{m_{3}%
}(x^{-})^{m_{-}}\big ]_{x\hspace{0.01in}=\hspace{0.01in}0}=\nonumber\\
=\delta_{m_{-},\hspace{0.01in}n_{-}}\delta_{m_{3},\hspace{0.01in}n_{3}}%
\delta_{m_{+},\hspace{0.01in}n_{+}}\delta_{m_{0},\hspace{0.01in}n_{0}}%
\,m_{0}!\,[[\hspace{0.01in}m_{+}]]_{q^{4}}!\,[[\hspace{0.01in}m_{3}]]_{q^{2}%
}!\,[[\hspace{0.01in}m_{-}]]_{q^{4}}!. \label{DuaPaaExtQuaSpa}%
\end{gather}
If we calculate dual pairings using right-actions of partial derivatives, we
obtain:%
\begin{gather}
\big \langle t^{\hspace{0.01in}m_{0}}(x^{+})^{m_{+}}(x^{3})^{m_{3}}%
(x^{-})^{m_{-}},(\text{i}E)^{n_{0}}(\text{i}^{-1}p_{-})^{n_{-}}(\text{i}%
^{-1}p_{3})^{n_{3}}(\text{i}^{-1}p_{+})^{n_{+}}\big \rangle_{x,\hspace
{0.01in}\bar{p}}=\nonumber\\
\big [t^{\hspace{0.01in}m_{0}}(x^{+})^{m_{+}}(x^{3})^{m_{3}}(x^{-})^{m_{-}%
}\,\bar{\triangleleft}\,(-\hspace{0.01in}\partial_{0})^{n_{0}}(-\hspace
{0.01in}\partial_{-})^{n_{-}}(-\hspace{0.01in}\partial_{3})^{n_{3}}%
(-\hspace{0.01in}\partial_{+})^{n_{+}}\big ]_{x\hspace{0.01in}=\hspace
{0.01in}0}=\nonumber\\
=\delta_{m_{-},\hspace{0.01in}n_{-}}\delta_{m_{3},\hspace{0.01in}n_{3}}%
\delta_{m_{+},\hspace{0.01in}n_{+}}\delta_{m_{0},\hspace{0.01in}n_{0}}%
\,m_{0}!\,[[\hspace{0.01in}m_{+}]]_{q^{4}}!\,[[\hspace{0.01in}m_{3}]]_{q^{2}%
}!\,[[\hspace{0.01in}m_{-}]]_{q^{4}}!. \label{DuaPaaExtQuaSpaRec}%
\end{gather}
From Eq.~(\ref{DuaPaaExtQuaSpa}) or Eq.~(\ref{DuaPaaExtQuaSpaRec}), we can
read off the elements being dual to the monomials $t^{\hspace{0.01in}m_{0}%
}(x^{+})^{m_{+}}(x^{3})^{m_{3}}(x^{-})^{m_{-}}$. This way,
Eq.~(\ref{GenForQExp}) enables us to write down expressions for $q$%
-exponentials of the three-dimensional $q$-deformed Euclidean space with a
time element. Concretely, we have%
\begin{align}
\exp_{q}(x|\text{i}p)  &  =\exp(\text{i}^{-1}tE)\hspace{0.01in}\exp
_{q}(\mathbf{x}|\text{i}\mathbf{p}),\nonumber\\
\exp_{q}(\text{i}^{-1}p\hspace{0.01in}|x)  &  =\exp(\text{i}Et)\hspace
{0.01in}\exp_{q}(\text{i}^{-1}\mathbf{p}\hspace{0.01in}|\mathbf{x}),
\label{ExpQuaDreExtEuk1}%
\end{align}
with the three-dimensional $q$-exponentials (also see Ref.~\cite{Wachter:2004ExpA})%
\begin{align}
\exp_{q}(\mathbf{x}|\text{i}\mathbf{p})  &  =\sum_{\underline{n}%
\,=\,0}^{\infty}\frac{(q\hspace{0.01in}x^{+})^{n_{+}}(x^{3})^{n_{3}}%
(q^{-1}x^{-})^{n_{-}}(\text{i}^{-1}p^{+})^{n_{-}}(\text{i}p^{3})^{n_{3}%
}(\text{i}^{-1}p^{-})^{n_{+}}}{[[\hspace{0.01in}n_{+}]]_{q^{4}}!\,[[\hspace
{0.01in}n_{3}]]_{q^{2}}!\,[[\hspace{0.01in}n_{-}]]_{q^{4}}!},\nonumber\\
\exp_{q}(\text{i}^{-1}\mathbf{p}|\mathbf{x})  &  =\sum_{\underline{n}%
\,=\,0}^{\infty}\frac{(\text{i}p^{+})^{n_{+}}(\text{i}^{-1}p^{3})^{n_{3}%
}(\text{i}p^{-})^{n_{-}}(q^{-1}x^{+})^{n_{-}}(x^{3})^{n_{3}}(q\hspace
{0.01in}x^{-})^{n_{+}}}{[[\hspace{0.01in}n_{+}]]_{q^{4}}!\,[[\hspace
{0.01in}n_{3}]]_{q^{2}}!\,[[\hspace{0.01in}n_{-}]]_{q^{4}}} \label{ExpEukExp}%
\end{align}
and the time-dependent phase factors%
\begin{equation}
\exp(\text{i}^{-1}t\otimes E)=\sum_{k\hspace{0.01in}=\hspace{0.01in}0}%
^{\infty}\frac{(\text{i}^{-1}t)^{k}\otimes E^{k}}{k!},\qquad\exp
(\text{i}E\otimes t)=\sum_{k\hspace{0.01in}=\hspace{0.01in}0}^{\infty}%
\frac{(\text{i}E)^{k}\otimes t^{k}}{k!}.
\end{equation}
Note that $E$ is a commutative parameter which we can interpret as energy.

\section{Time evolution operator\label{KapZeiEntOpeN}}

The $q$-exponentials of the quantum space $\mathcal{\mathbb{R}}_{q}^{3}$
provide us with an operator that generates spatial displacements. We obtain
this operator from the expressions for the $q$-exponentials given in
Eq.~(\ref{ExpEukExp}) of the previous chapter if we replace the momentum
coordinates $p^{A}$ by the derivative operators i$^{-1}\partial^{A}$
\cite{Carnovale:1999,Majid:1993ud,Wachter:2007A}:%
\begin{align}
\exp_{q}(x|\partial_{y})\triangleright g(\hspace{0.01in}y)  &  =g(x\,\bar
{\oplus}\,y),\nonumber\\
g(\hspace{0.01in}y)\,\bar{\triangleleft}\,\exp_{q}(-\hspace{0.01in}%
\partial_{y}|\hspace{0.01in}x)  &  =g(\hspace{0.01in}y\,\bar{\oplus}\,x).
\label{q-TayN}%
\end{align}

We recall that $q$-translations and $q$-inversions are realizations of braided
co-products and braided antipodes, respectively [cf. Eqs.~(\ref{KonReaBraCop})
and (\ref{qInvDef}) of Chap.~\ref{KapTra}]. The braided co-products and
braided antipodes satisfy the axioms (also see Ref.~\cite{Majid:1996kd})%
\begin{align}
m\circ(\underline{S}\otimes\operatorname*{id})\circ\underline{\Delta}  &
=m\circ(\operatorname*{id}\otimes\,\underline{S}\hspace{0.01in})\circ
\underline{\Delta}=\underline{\varepsilon},\nonumber\\
m\circ(\underline{\bar{S}}\otimes\operatorname*{id})\circ\underline
{\bar{\Delta}}  &  =m\circ(\operatorname*{id}\otimes\,\underline{\bar{S}%
}\hspace{0.01in})\circ\underline{\bar{\Delta}}=\underline{\bar{\varepsilon}},
\label{HopfVerAnfN}%
\end{align}
and%
\begin{align}
(\operatorname*{id}\otimes\,\underline{\varepsilon})\circ\underline{\Delta}
&  =\operatorname*{id}=(\underline{\varepsilon}\otimes\operatorname*{id}%
)\circ\underline{\Delta},\nonumber\\
(\operatorname*{id}\otimes\,\underline{\bar{\varepsilon}})\circ\underline
{\bar{\Delta}}  &  =\operatorname*{id}=(\underline{\bar{\varepsilon}}%
\otimes\operatorname*{id})\circ\underline{\bar{\Delta}}. \label{HopfAxi2}%
\end{align}
In the identities above, we denote the operation of multiplication on the
braided Hopf algebra by $m$. The co-units $\underline{\varepsilon}%
,\underline{\bar{\varepsilon}}$ of the two braided Hopf structures are both
linear mappings that vanish on the coordinate generators:%
\begin{equation}
\varepsilon(X^{i})=\underline{\bar{\varepsilon}}(X^{i})=0.
\end{equation}
For this reason, we can realize the co-units $\underline{\varepsilon}$ and
$\underline{\bar{\varepsilon}}$ on a commutative coordinate algebra as
follows:%
\begin{equation}
\underline{\varepsilon}(\mathcal{W}(f))=\underline{\bar{\varepsilon}%
}(\mathcal{W}(f))=\left.  f(x)\right\vert _{x\hspace{0.01in}=\hspace{0.01in}%
0}=f(0). \label{ReaVerZopNeuEleKomAlg}%
\end{equation}
Now, we are in a position to translate the Hopf algebra axioms in
Eqs.~(\ref{HopfVerAnfN}) and (\ref{HopfAxi2}) into corresponding rules for
$q$-translations and $q$-inversions \cite{Wachter:2007A}, i.~e.%
\begin{align}
f((\ominus\,x)\oplus x)  &  =f(x\oplus(\ominus\,x))=f(0),\nonumber\\
f((\bar{\ominus}\,x)\,\bar{\oplus}\,x)  &  =f(x\,\bar{\oplus}\,(\bar{\ominus
}\,x))=f(0), \label{qAddN}%
\end{align}
and%
\begin{align}
f(x\oplus y)|_{y\hspace{0.01in}=\hspace{0.01in}0}  &  =f(x)=f(y\oplus
x)|_{y\hspace{0.01in}=\hspace{0.01in}0},\nonumber\\
f(x\,\bar{\oplus}\,y)|_{y\hspace{0.01in}=\hspace{0.01in}0}  &  =f(x)=f(y\,\bar
{\oplus}\,x)|_{y\hspace{0.01in}=\hspace{0.01in}0}. \label{qNeuEle}%
\end{align}

With the help of the rules written down in Eqs.~(\ref{qAddN}) and
(\ref{qNeuEle}), the identities in Eq.~(\ref{q-TayN}) imply%
\[
\exp_{q}(x\,\bar{\oplus}\,(\bar{\ominus}\,y)|\hspace{0.01in}\partial
_{y})\triangleright g(\hspace{0.01in}y)=g(x\,\bar{\oplus}\,(\bar{\ominus
}\,y)\,\bar{\oplus}\,y)=g(x)
\]
and%
\begin{equation}
\lbrack\hspace{0.01in}\exp_{q}(x|\partial_{y})\triangleright g(\hspace
{0.01in}y)]_{\hspace{0.01in}y\hspace{0.01in}=\hspace{0.01in}0}=[\hspace
{0.01in}g(x\,\bar{\oplus}\,y)]_{\hspace{0.01in}y\hspace{0.01in}=\hspace
{0.01in}0}=g(x).
\end{equation}
We can combine the above results as follows:%
\begin{equation}
\exp_{q}(x\,\bar{\oplus}\,(\bar{\ominus}\,y)|\hspace{0.01in}\partial
_{y})\triangleright g(\hspace{0.01in}y)=[\hspace{0.01in}\exp_{q}%
(x|\partial_{y})\triangleright g(\hspace{0.01in}y)]_{\hspace{0.01in}%
y\hspace{0.01in}=\hspace{0.01in}0}=g(x). \label{qTayLink}%
\end{equation}
Similar identities hold for right-actions:%
\begin{equation}
g(\hspace{0.01in}y)\,\bar{\triangleleft}\,\exp_{q}(-\hspace{0.01in}%
\partial_{y}|(\bar{\ominus}\,y)\,\bar{\oplus}\,x)=[\hspace{0.01in}%
g(y)\,\bar{\triangleleft}\,\exp_{q}(-\hspace{0.01in}\partial_{y}%
|\hspace{0.01in}x)]_{\hspace{0.01in}y\hspace{0.01in}=\hspace{0.01in}0}=g(x).
\label{qTayRec}%
\end{equation}

If we apply the results of Eq.~(\ref{qTayLink}) or Eq.~(\ref{qTayRec}) to the
$q$-exponentials\ given in Eq.~(\ref{ExpQuaDreExtEuk1}) of the last chapter,
we obtain operators for displacing functions in space and time, i.~e.%
\begin{equation}
\exp((t-t^{\hspace{0.01in}\prime})\otimes\partial_{t^{\prime}})\hspace
{0.01in}\exp_{q}(\mathbf{x}\,\bar{\oplus}\,(\bar{\ominus}\,\mathbf{y}%
)|\hspace{0.01in}\partial_{\mathbf{y}})\triangleright g(\mathbf{y}%
,\hspace{0.01in}t^{\hspace{0.01in}\prime})=g(\mathbf{x},t)
\end{equation}
or%
\begin{equation}
g(\hspace{0.01in}\mathbf{y},t^{\hspace{0.01in}\prime})\,\bar{\triangleleft
}\,\exp_{q}(-\hspace{0.01in}\partial_{\mathbf{y}}|(\bar{\ominus}%
\,\mathbf{y})\,\bar{\oplus}\,\mathbf{x})\hspace{0.01in}\exp(\partial
_{t^{\prime}}\otimes(t^{\hspace{0.01in}\prime}\hspace{-0.01in}%
-t))=g(\mathbf{x},t).
\end{equation}
Since the time coordinate is independent of the space coordinates, we can
perform time-displacements independently of space-displacements. For this
reason, we also have:%
\begin{align}
\exp((t-t^{\hspace{0.01in}\prime})\otimes\partial_{t^{\prime}})\triangleright
g(\mathbf{x},t^{\hspace{0.01in}\prime})  &  =[\hspace{0.01in}\exp
(t\otimes\partial_{t^{\prime}})\triangleright g(\mathbf{x},t^{\hspace
{0.01in}\prime})]_{t^{\prime}=\hspace{0.01in}0}=g(\mathbf{x},t),\nonumber\\
g(\mathbf{x},t^{\hspace{0.01in}\prime})\triangleleft\exp(\partial_{t^{\prime}%
}\otimes(t^{\hspace{0.01in}\prime}\hspace{-0.01in}-t))  &  =[\hspace
{0.01in}g(\mathbf{x},\hspace{0.01in}t^{\hspace{0.01in}\prime})\triangleleft
\hspace{0.01in}\exp(\partial_{t^{\prime}}\otimes(-\hspace{0.01in}%
t))]_{t^{\prime}=\hspace{0.01in}0}=g(\mathbf{x},t).
\end{align}

In quantum theory, we can calculate the time evolution of a wave function from
its values at all points in space at a given time. To see this, we recall the
following facts. The time evolution operator is obtained from the operator for
a time shift if we replace the time derivative with the Hamilton operator. The
Hamilton operator, however, acts on the spatial coordinates, only.

These facts also hold for the $q$-deformed Euclidean space. Let $\phi
(\mathbf{x},t)$ be a $q$-deformed wave function describing the quantum state
of a system with Hamilton operator $H$. We can calculate $\phi(\mathbf{x},t)$
from $\phi(\mathbf{x},t^{\hspace{0.01in}\prime}\hspace{-0.01in}=0)$ using the
identities%
\begin{align}
\phi(\mathbf{x},t)  &  =\hspace{0.01in}\mathcal{U}(t,t^{\hspace{0.01in}\prime
}\hspace{-0.01in}=0)\triangleright\phi(\mathbf{x},t^{\hspace{0.01in}\prime
}\hspace{-0.01in}=0),\nonumber\\
\bar{\phi}(\mathbf{x},t)  &  =\bar{\phi}(\mathbf{x},t^{\hspace{0.01in}\prime
}\hspace{-0.01in}=0)\triangleleft\hspace{0.01in}\overline{\mathcal{U}%
}(t,t^{\hspace{0.01in}\prime}\hspace{-0.01in}=0) \label{TimEvoId1}%
\end{align}
if the time evolution operators are given by the following expressions:%
\begin{align}
\mathcal{U}(t,t^{\hspace{0.01in}\prime}\hspace{-0.01in}  &  =0)=\exp
(-\hspace{0.01in}t\otimes\text{i}H),\nonumber\\
\overline{\mathcal{U}}(t,t^{\hspace{0.01in}\prime}\hspace{-0.01in}  &
=0)=\exp(\text{i}H\otimes t). \label{ZeiEntQuaEukDre}%
\end{align}
A look at Eq.~(\ref{ZeiEntQuaEukDre}) shows that the time development
operators $\mathcal{U}$ and $\overline{\mathcal{U}}$ of the $q$-deformed
Euclidean space have the same form as in the undeformed case. Thus, both time
development operators show the same properties as in the undeformed case
\cite{Sakurai:1994}.

We can immediately specify the inverse of both time evolution operators, i.~e.%
\begin{align}
&  \mathcal{U}^{-1}(t,t^{\hspace{0.01in}\prime}\hspace{-0.01in}=0)=\hspace
{0.01in}\mathcal{U}(-\hspace{0.01in}t,t^{\hspace{0.01in}\prime}\hspace
{-0.01in}=0)=\exp(t\otimes\text{i}H),\nonumber\\[0.03in]
&  \overline{\mathcal{U}}{}^{-1}(t,t^{\hspace{0.01in}\prime}\hspace
{-0.01in}=0)=\hspace{0.01in}\overline{\mathcal{U}}(-\hspace{0.01in}%
t,t^{\hspace{0.01in}\prime}\hspace{-0.01in}=0)=\exp(-\text{i}H\otimes t),
\end{align}
with%
\begin{align}
\mathcal{U}(t,t^{\hspace{0.01in}\prime}\hspace{-0.01in}=0)\,\mathcal{U}%
^{-1}(t,t^{\hspace{0.01in}\prime}\hspace{-0.01in}=0)=\,\,  &  \mathcal{U}%
^{-1}(t,t^{\hspace{0.01in}\prime}\hspace{-0.01in}=0)\,\mathcal{U}%
(t,t^{\hspace{0.01in}\prime}\hspace{-0.01in}=0)=1,\nonumber\\[0.03in]
\overline{\mathcal{U}}{}^{-1}(t,t^{\hspace{0.01in}\prime}\hspace
{-0.01in}=0)\,\overline{\mathcal{U}}(t,t^{\hspace{0.01in}\prime}%
\hspace{-0.01in}=0)=\,\,  &  \overline{\mathcal{U}}(t,t^{\hspace{0.01in}%
\prime}\hspace{-0.01in}=0)\,\overline{\mathcal{U}}{}^{-1}(t,t^{\hspace
{0.01in}\prime}\hspace{-0.01in}=0)=1. \label{InvZeiDefGle}%
\end{align}
The inverse of each time evolution operator transforms wave functions into the
opposite time direction:%
\begin{align}
\phi(\mathbf{x},-\hspace{0.01in}t)  &  =\hspace{0.01in}\mathcal{U}%
^{-1}(t,t^{\hspace{0.01in}\prime}\hspace{-0.01in}=0)\triangleright
\phi(\mathbf{x},t^{\hspace{0.01in}\prime}\hspace{-0.01in}=0),\nonumber\\
\bar{\phi}(\mathbf{x},-\hspace{0.01in}t)  &  =\bar{\phi}(\mathbf{x}%
,t^{\hspace{0.01in}\prime}\hspace{-0.01in}=0)\triangleleft\,\overline
{\mathcal{U}}{}^{-1}(t,t^{\hspace{0.01in}\prime}\hspace{-0.01in}=0).
\end{align}
If we apply the inverse time evolution operators to Eq.~(\ref{TimEvoId1}) and
take into account Eq.~(\ref{InvZeiDefGle}), we also get:%
\begin{align}
\phi(\mathbf{x},t^{\hspace{0.01in}\prime}\hspace{-0.01in}  &  =0)=\hspace
{0.01in}\mathcal{U}^{-1}(t,t^{\hspace{0.01in}\prime}\hspace{-0.01in}%
=0)\triangleright\phi(\mathbf{x},t),\nonumber\\
\bar{\phi}(\mathbf{x},t^{\hspace{0.01in}\prime}\hspace{-0.01in}  &
=0)=\bar{\phi}(\mathbf{x},t)\triangleleft\,\overline{\mathcal{U}}{}%
^{-1}(t,t^{\hspace{0.01in}\prime}\hspace{-0.01in}=0). \label{Tim2N}%
\end{align}

The time evolution operators in Eq.~(\ref{ZeiEntQuaEukDre}) shift the wave
functions from the time-zero point. We can eliminate this restriction by
generalizing the time evolution operators in the following way:%
\begin{align}
\mathcal{U}(t,t^{\hspace{0.01in}\prime})  &  =\hspace{0.01in}\mathcal{U}%
(t,t^{\hspace{0.01in}\prime\prime}\hspace{-0.01in}=0)\,\mathcal{U}%
^{-1}(t^{\hspace{0.01in}\prime},t^{\hspace{0.01in}\prime\prime}\hspace
{-0.01in}=0)=\exp(-(t-t^{\hspace{0.01in}\prime})\otimes\text{i}%
H),\nonumber\\[0.03in]
\overline{\mathcal{U}}(t,t^{\hspace{0.01in}\prime})  &  =\hspace
{0.01in}\overline{\mathcal{U}}{}^{-1}(t^{\hspace{0.01in}\prime},t^{\hspace
{0.01in}\prime\prime}\hspace{-0.01in}=0)\,\overline{\mathcal{U}}%
(t,t^{\hspace{0.01in}\prime\prime}\hspace{-0.01in}=0)=\exp(-\text{i}%
H\otimes(t^{\hspace{0.01in}\prime}\hspace{-0.01in}-t)). \label{GenTimEvo2}%
\end{align}
Applying Eqs.~(\ref{TimEvoId1}), (\ref{Tim2N}), and (\ref{GenTimEvo2}), we can
see that the operators above transform wave functions from time $t^{\hspace
{0.01in}\prime}$ to time $t$:%
\begin{align}
\phi(\mathbf{x},t)  &  =\hspace{0.01in}\mathcal{U}(t,t^{\hspace{0.01in}\prime
})\triangleright\phi(\mathbf{x},t^{\hspace{0.01in}\prime}),\nonumber\\
\bar{\phi}(\mathbf{x},t)  &  =\bar{\phi}(\mathbf{x},t^{\hspace{0.01in}\prime
})\triangleleft\,\overline{\mathcal{U}}(t,t^{\hspace{0.01in}\prime}).
\label{TimDisp}%
\end{align}

The general time evolution operators in Eq.~(\ref{GenTimEvo2}) have an inverse
again, i.~e. there are operators $\mathcal{U}^{-1}(t,t^{\hspace{0.01in}\prime
})$ and $\overline{\mathcal{U}}{}^{-1}(t,t^{\hspace{0.01in}\prime})$ with%
\begin{align}
\mathcal{U}(t,t^{\hspace{0.01in}\prime})\,\mathcal{U}^{-1}(t,t^{\hspace
{0.01in}\prime})  &  =\hspace{0.01in}\mathcal{U}^{-1}(t,t^{\hspace
{0.01in}\prime})\,\mathcal{U}(t,t^{\hspace{0.01in}\prime}%
)=1,\nonumber\\[0.03in]
\overline{\mathcal{U}}(t,t^{\hspace{0.01in}\prime})\,\overline{\mathcal{U}}%
{}^{-1}(t,t^{\hspace{0.01in}\prime})  &  =\hspace{0.01in}\overline
{\mathcal{U}}{}^{-1}(t,t^{\hspace{0.01in}\prime})\,\overline{\mathcal{U}%
}(t,t^{\hspace{0.01in}\prime})=1.
\end{align}
These identities are satisfied by the following expressions:%
\begin{align}
\mathcal{U}^{-1}(t,t^{\hspace{0.01in}\prime})  &  =\hspace{0.01in}%
\mathcal{U}(t^{\hspace{0.01in}\prime},t^{\hspace{0.01in}\prime\prime}%
\hspace{-0.01in}=0)\,\mathcal{U}^{-1}(t,t^{\hspace{0.01in}\prime\prime}%
\hspace{-0.01in}=0)=\hspace{0.01in}\mathcal{U}(t^{\hspace{0.01in}\prime
},t)\nonumber\\
&  =\hspace{0.01in}\mathcal{U}(-\hspace{0.01in}t,-\hspace{0.01in}t^{\prime
})=\exp((t-t^{\hspace{0.01in}\prime})\otimes\text{i}H),\\[0.1in]
\overline{\mathcal{U}}{}^{-1}(t,t^{\hspace{0.01in}\prime})  &  =\hspace
{0.01in}\overline{\mathcal{U}}{}^{-1}(t,t^{\hspace{0.01in}\prime\prime}%
\hspace{-0.01in}=0)\,\overline{\mathcal{U}}(t^{\hspace{0.01in}\prime
},t^{\hspace{0.01in}\prime\prime}\hspace{-0.01in}=0)=\hspace{0.01in}%
\overline{\mathcal{U}}(t^{\hspace{0.01in}\prime},t)\nonumber\\
&  =\hspace{0.01in}\overline{\mathcal{U}}(-\hspace{0.01in}t,-\hspace
{0.01in}t^{\hspace{0.01in}\prime})=\exp(\text{i}H\otimes(t^{\hspace
{0.01in}\prime}-t)).
\end{align}

The operator $\mathcal{U}(t,t^{\hspace{0.01in}\prime})$ describes the
evolution from time $t^{\hspace{0.01in}\prime}$ to time $t$, and the inverse
operator $\mathcal{U}^{-1}(t,t^{\hspace{0.01in}\prime})$ turns around this
evolution. The same holds for the operators $\overline{\mathcal{U}%
}(t,t^{\hspace{0.01in}\prime})$ and $\overline{\mathcal{U}}{}^{-1}%
(t,t^{\hspace{0.01in}\prime})$. Thus, we have in analogy to Eq.~(\ref{TimDisp}%
):%
\begin{align}
\phi(\mathbf{x},t^{\hspace{0.01in}\prime})  &  =\hspace{0.01in}\mathcal{U}%
^{-1}(t,t^{\hspace{0.01in}\prime})\triangleright\phi(\mathbf{x},t),\nonumber\\
\bar{\phi}(\mathbf{x},t^{\hspace{0.01in}\prime})  &  =\bar{\phi}%
(\mathbf{x},t)\triangleleft\,\overline{\mathcal{U}}{}^{-1}(t,t^{\hspace
{0.01in}\prime}).
\end{align}

The time evolution operators defined in Eq.~(\ref{GenTimEvo2}) satisfy the
\textit{principle of causality}. Accordingly, we can write the successive
action of two different time evolution operators as the action of one single
time evolution operator. Concretely, we have%
\begin{align}
\mathcal{U}(t,t^{\hspace{0.01in}\prime\prime})\,\mathcal{U}(t^{\hspace
{0.01in}\prime\prime},t^{\hspace{0.01in}\prime})  &  =\hspace{0.01in}%
\mathcal{U}(t,0)\,\mathcal{U}^{-1}(t^{\hspace{0.01in}\prime\prime
},0)\,\mathcal{U}(t^{\hspace{0.01in}\prime\prime},0)\,\mathcal{U}%
^{-1}(t^{\hspace{0.01in}\prime},0)\nonumber\\
&  =\hspace{0.01in}\mathcal{U}(t,0)\,\mathcal{U}^{-1}(t^{\hspace{0.01in}%
\prime},0)=\hspace{0.01in}\mathcal{U}(t,t^{\hspace{0.01in}\prime})
\end{align}
and%
\begin{align}
\overline{\mathcal{U}}(t^{\hspace{0.01in}\prime\prime},t^{\hspace
{0.01in}\prime})\,\overline{\mathcal{U}}(t,t^{\hspace{0.01in}\prime\prime})
&  =\hspace{0.01in}\overline{\mathcal{U}}{}^{-1}(t^{\hspace{0.01in}\prime
},0)\,\overline{\mathcal{U}}(t^{\hspace{0.01in}\prime\prime},0)\,\overline
{\mathcal{U}}{}^{-1}(t^{\hspace{0.01in}\prime\prime},0)\,\overline
{\mathcal{U}}(t,0)\nonumber\\
&  =\hspace{0.01in}\overline{\mathcal{U}}{}^{-1}(t^{\hspace{0.01in}\prime
},0)\,\overline{\mathcal{U}}(t,0)=\hspace{0.01in}\overline{\mathcal{U}%
}(t,t^{\hspace{0.01in}\prime}).
\end{align}

Recall that the time element has trivial braiding properties and that the
Hamilton operator acts on the space coordinates, only. For this reason, the
Hamilton operator commutes with the time variable, and we can drop the symbol
for the tensor product in the expressions for the time evolution operator:%
\begin{equation}
\exp(t\otimes\text{i}H)=\exp(\text{i}H\otimes t)=\exp(\text{i}Ht).
\end{equation}
Accordingly, we can also write:%
\begin{equation}
\mathcal{U}(t,t^{\hspace{0.01in}\prime})=\exp(-\text{i}H(t-t^{\hspace
{0.01in}\prime})),\qquad\overline{\mathcal{U}}(t,t^{\hspace{0.01in}\prime
})=\exp(\text{i}H(t-t^{\hspace{0.01in}\prime})). \label{TimeEvolClas}%
\end{equation}
With this result we can make the following identifications:%
\begin{equation}
\mathcal{U}(t,t^{\hspace{0.01in}\prime})=\hspace{0.01in}\overline{\mathcal{U}%
}(-\hspace{0.01in}t,-\hspace{0.01in}t^{\hspace{0.01in}\prime})=\hspace
{0.01in}\mathcal{U}^{-1}(-\hspace{0.01in}t,-\hspace{0.01in}t^{\hspace
{0.01in}\prime})=\hspace{0.01in}\overline{\mathcal{U}}{}^{-1}(t,t^{\hspace
{0.01in}\prime}).
\end{equation}
From Eq.~(\ref{TimeEvolClas}) also follows that the time evolution operators
are unitary since the Hamilton operator is Hermitian:%
\begin{equation}
\mathcal{U}^{-1}(t,t^{\hspace{0.01in}\prime})=\hspace{0.01in}\mathcal{U}%
(t^{\hspace{0.01in}\prime},t)=\hspace{0.01in}\overline{\mathcal{U}%
}(t,t^{\hspace{0.01in}\prime})=\hspace{0.01in}\mathcal{U}^{\dag}%
(t,t^{\hspace{0.01in}\prime}). \label{UntZeitEntOpeDreQuaEuk}%
\end{equation}

\section{Schr\"{o}dinger picture and Heisenberg picture\label{KapSchHeiBil}}

As explained in the last chapter, the time evolution operator of the
three-dimensional $q$-deformed Euclidean space takes on the same form as in
the undeformed case. This fact enables us to apply the well-known methods for
describing the time evolution of a quantum mechanical system to the
three-dimensional $q$-deformed Euclidean space.

First, we derive a differential equation for the time evolution operator
$\mathcal{U}(t,t^{\hspace{0.01in}\prime})$:%
\begin{align}
\partial_{t}\triangleright\mathcal{U}(t,t^{\hspace{0.01in}\prime})  &
=\partial_{t}\triangleright\mathcal{U}(t,0)\,\mathcal{U}^{-1}(t^{\hspace
{0.01in}\prime},0)=\partial_{t}\triangleright\exp(-\hspace{0.01in}%
t\otimes\text{i}H)\,\mathcal{U}^{-1}(t^{\hspace{0.01in}\prime},0)\nonumber\\
&  =\exp(-\hspace{0.01in}t\otimes\text{i}H)\hspace{0.01in}(-\text{i}%
H)\,\mathcal{U}^{-1}(t^{\hspace{0.01in}\prime},0)=-\text{i}H\,\mathcal{U}%
(t,0)\,\mathcal{U}^{-1}(t^{\hspace{0.01in}\prime},0)\nonumber\\
&  =-\text{i}H\,\mathcal{U}(t,t^{\hspace{0.01in}\prime}).
\label{BerSchGleZeiEnt1}%
\end{align}
In the above calculation, we have made use of Eqs.$~$(\ref{ZeiEntQuaEukDre})
and (\ref{GenTimEvo2}) of the previous chapter. We can proceed in the same way
for the time evolution operator corresponding to right-actions:%
\begin{align}
\overline{\mathcal{U}}(t,t^{\hspace{0.01in}\prime})\triangleleft\partial_{t}
&  =\hspace{0.01in}\overline{\mathcal{U}}{}^{-1}(t^{\hspace{0.01in}\prime
},0)\,\overline{\mathcal{U}}(t,0)\triangleleft\partial_{t}=\hspace
{0.01in}\overline{\mathcal{U}}{}^{-1}(t^{\hspace{0.01in}\prime},0)\hspace
{0.01in}\exp(\text{i}H\otimes t)\triangleleft\partial_{t}\nonumber\\
&  =\hspace{0.01in}\overline{\mathcal{U}}{}^{-1}(t^{\hspace{0.01in}\prime
},0)\hspace{0.01in}(-\text{i}H)\hspace{0.01in}\exp(\text{i}H\otimes
t)=\hspace{0.01in}\overline{\mathcal{U}}{}^{-1}(t^{\hspace{0.01in}\prime
},0)\,\overline{\mathcal{U}}(t,0)\hspace{0.01in}(-\text{i}H)\nonumber\\
&  =\hspace{0.01in}\overline{\mathcal{U}}(t,t^{\hspace{0.01in}\prime}%
)\hspace{0.01in}(-\text{i}H). \label{BerSchGleZeiEnt2}%
\end{align}
The calculations in Eqs.~(\ref{BerSchGleZeiEnt1}) and (\ref{BerSchGleZeiEnt2})
lead us to the so-called \textit{Schr\"{o}dinger equations of the time
evolution operators}. If we take into account that the operator
representations of $\partial_{0}$ and $\hat{\partial}_{0}$ are nothing else
but the usual time derivative [cf. Eqs.~(\ref{OpeDarZeiAblExtQuaEuk}) and
(\ref{OpeDarZeiAblExtQuaEukKon}) of Chap.~\ref{ParAblKapAna}], we can write
these Schr\"{o}dinger equations as follows:%
\begin{align}
\text{i}\partial_{0}\triangleright\mathcal{U}(t,t^{\hspace{0.01in}\prime})  &
=H\,\mathcal{U}(t,t^{\hspace{0.01in}\prime}), & \mathcal{U}(t^{\hspace
{0.01in}\prime},t)\,\bar{\triangleleft}\,\partial_{0}\text{i}  &
=\mathcal{U}(t^{\hspace{0.01in}\prime},t)\hspace{0.01in}H,\nonumber\\
\text{i}\hat{\partial}_{0}\,\bar{\triangleright}\,\hspace{0.01in}%
\mathcal{\hat{U}}(t,t^{\hspace{0.01in}\prime})  &  =\hat{H}\,\mathcal{\hat{U}%
}(t,t^{\hspace{0.01in}\prime}), & \mathcal{\hat{U}}(t^{\hspace{0.01in}\prime
},t)\triangleleft\hat{\partial}_{0}\text{i}  &  =\mathcal{\hat{U}}%
(t^{\hspace{0.01in}\prime},t)\hspace{0.01in}\hat{H}. \label{SchoEq1}%
\end{align}
Note that the Hamilton operators $H$ and $\hat{H}$ depend on $\partial^{A}$
and $\hat{\partial}^{A}$, respectively [also see Eqs.~(\ref{UnkOpeDarAbl}) and
(\ref{KonOpeDarAbl}) of Chap.~\ref{ParAblKapAna}].

The two Schr\"{o}dinger equations of $\mathcal{U}$ have their equivalent in
the integral equations%
\begin{align}
\mathcal{U}(t,t^{\hspace{0.01in}\prime})  &  =1-\text{i}\int
\nolimits_{t^{\prime}}^{t}\text{d}t^{\hspace{0.01in}\prime\prime
}\,H(t^{\hspace{0.01in}\prime\prime})\,\mathcal{U}(t^{\hspace{0.01in}%
\prime\prime},t^{\hspace{0.01in}\prime}),\nonumber\\
\mathcal{U}(t^{\hspace{0.01in}\prime},t)  &  =1+\text{i}\int
\nolimits_{t^{\prime}}^{t}\text{d}t^{\hspace{0.01in}\prime\prime}%
\,\mathcal{U}(t^{\hspace{0.01in}\prime},t^{\hspace{0.01in}\prime\prime
})\,H(t^{\hspace{0.01in}\prime\prime}) \label{IntGleZeiEntOpeSchr}%
\end{align}
if we assume the following constraint:%
\begin{equation}
\mathcal{U}(t,t)=1.
\end{equation}
The above integral equations have formal solutions%
\begin{equation}
\mathcal{U}(t,t^{\hspace{0.01in}\prime})=1+\sum_{n\hspace{0.01in}%
=\hspace{0.01in}1}^{\infty}\text{i}^{-n}\int\limits_{t^{\prime}}^{t}%
\text{d}t_{1}\int\limits_{t^{\prime}}^{t_{1}}\text{d}t_{2}\hspace
{0.01in}\ldots\hspace{-0.01in}\int\limits_{t^{\prime}}^{t_{n-1}}\text{d}%
t_{n}\,H(t_{1})\hspace{0.01in}H(t_{2})\ldots\hspace{0.01in}H(t_{n})
\label{IteEntZeiEntOpe}%
\end{equation}
and%
\[
\mathcal{U}(t^{\hspace{0.01in}\prime},t)=1+\sum_{n\hspace{0.01in}%
=\hspace{0.01in}1}^{\infty}\text{i}^{n}\int\limits_{t^{\prime}}^{t}%
\text{d}t_{1}\int\limits_{t^{\prime}}^{t_{1}}\text{d}t_{2}\hspace
{0.01in}\ldots\hspace{-0.01in}\int\limits_{t^{\prime}}^{t_{n-1}}\text{d}%
t_{n}\,H(t_{n})\hspace{0.01in}H(t_{n-1})\ldots\hspace{0.01in}H(t_{1}).
\]
For the operators $\mathcal{\hat{U}}$ and $\hat{H}$, similar formulas apply,
which we obtain by replacing $\mathcal{U}$ and $H$ with $\mathcal{\hat{U}}$
and $\hat{H}$, respectively.

In quantum theory, there are two ways of describing the time evolution of a
quantum system, namely the Schr\"{o}dinger picture and the Heisenberg picture.
In the \textit{Schr\"{o}dinger picture}, the time evolution is determined by
the time dependence of the wave functions, whereas the observables are usually
independent of time. We can derive the equations of motion for the wave
functions by using Eqs.~(\ref{TimDisp}) and (\ref{SchoEq1}):%
\begin{align}
\text{i}\partial_{0}\triangleright\phi(\mathbf{x},t)  &  =\text{i}\partial
_{t}\triangleright\mathcal{U}(t,t^{\hspace{0.01in}\prime})\triangleright
\phi(\mathbf{x},t^{\hspace{0.01in}\prime})\nonumber\\
&  =H\,\mathcal{U}(t,t^{\hspace{0.01in}\prime})\triangleright\phi
(\mathbf{x},t^{\hspace{0.01in}\prime})=H\triangleright\phi(\mathbf{x},t).
\label{SchrEqWav}%
\end{align}
Similarly, we get:%
\begin{align}
\phi(\mathbf{x},t)\triangleleft\hat{\partial}_{0}\text{i}  &  =\phi
(\mathbf{x},t^{\hspace{0.01in}\prime})\triangleleft\hspace{0.01in}%
\mathcal{\hat{U}}(t^{\hspace{0.01in}\prime},t)\triangleleft\partial
_{t}\text{i}\nonumber\\
&  =\phi(\mathbf{x},t^{\hspace{0.01in}\prime})\triangleleft\hspace
{0.01in}\mathcal{\hat{U}}(t^{\hspace{0.01in}\prime},t)\hspace{0.01in}\hat
{H}=\phi(\mathbf{x},t)\triangleleft\hat{H}.
\end{align}
We obtain further equations of motion by applying the following substitution
to the above identities:%
\begin{equation}
\partial_{0}\leftrightarrow\hat{\partial}_{0},\qquad\mathcal{U}\leftrightarrow
\mathcal{\hat{U}},\qquad H\leftrightarrow\hat{H},\qquad\triangleright
\leftrightarrow\bar{\triangleright},\qquad\triangleleft\leftrightarrow
\bar{\triangleleft}.
\end{equation}
Due to these substitution rules, we will restrict ourselves to the time
development operator $\mathcal{U}$ and the Hamilton operator $H$.

We show in the appendix that the following expression defines a $q$-deformed
scalar product of two time-dependent wave functions:%
\begin{equation}
\big \langle\phi(t)|\hspace{0.01in}\psi(t)\big \rangle_{q}=\int\text{d}%
_{q}^{3}x\,\overline{\phi(\mathbf{x},t)}\circledast\psi(\mathbf{x},t).
\label{DefSesq}%
\end{equation}
The time dependence of the scalar product results from the wave functions
only. We now show that the above scalar product does not change in time if the
time development operator is unitary:%
\begin{align}
\big \langle\phi(t)|\hspace{0.01in}\psi(t)\big \rangle_{q}  &  =\int
\text{d}_{q}^{3}x\,\overline{\phi(\mathbf{x},t)}\circledast\psi(\mathbf{x}%
,t)\nonumber\\
&  =\int\text{d}_{q}^{3}x\,[\hspace{0.01in}\overline{\phi(x,0)}\triangleleft
\mathcal{U}^{\dag}(t,0)]\circledast\lbrack\,\mathcal{U}(t,0)\triangleright
\psi(\mathbf{x},0)]\nonumber\\
&  =\int\text{d}_{q}^{3}x\,\overline{\phi(\mathbf{x},0)}\circledast
\lbrack\,\mathcal{U}^{-1}(t,0)\,\mathcal{U}(t,0)\triangleright\psi
(\mathbf{x},0)]\nonumber\\
&  =\int\text{d}_{q}^{3}x\,\overline{\phi(\mathbf{x},0)}\circledast
\psi(\mathbf{x},0)=\big \langle\phi(0)|\hspace{0.01in}\psi(0)\big \rangle_{q}.
\label{CalTimDevSes}%
\end{align}
In the above calculation, we made use of Eqs.~(\ref{TimEvoId1}) and
(\ref{UntZeitEntOpeDreQuaEuk}) of the previous chapter. In addition to this,
we took into account that the operators $\mathcal{U}(t,0)$ and $\mathcal{U}%
^{-1}(t,0)$ depend on the partial derivatives $\partial^{A}$ for which we have
the following $q$-analog of Stokes' theorem \cite{Wachter:2007A}:%
\begin{equation}
\int_{-\infty}^{+\infty}\text{d}_{q}^{3}x\,(f\triangleleft\partial
^{A})\circledast g=\int_{-\infty}^{+\infty}\text{d}_{q}^{3}x\,f\circledast
(\partial^{A}\triangleright g). \label{PatIntUneRaumInt}%
\end{equation}
The result of Eq.~(\ref{CalTimDevSes}) implies that the normalization of a
wave function does not change over time:%
\begin{equation}
\big \langle\phi(t)|\hspace{0.01in}\phi(t)\big \rangle_{q}=1. \label{NorBed1}%
\end{equation}

Next, we examine the time dependence of matrix elements of an observable,
which in the following is denoted by $\hat{O}$. With similar considerations as
in Eq.~(\ref{CalTimDevSes}), we get:%
\begin{align}
\big \langle\phi(t)|\hspace{0.01in}\hat{O}\triangleright\psi
(t)\big \rangle_{q}  &  =\int\text{d}_{q}^{3}x\,\overline{\phi(\mathbf{x}%
,t)}\circledast(\hat{O}\triangleright\psi(\mathbf{x},t))\nonumber\\
&  =\int\text{d}_{q}^{3}x\hspace{0.03in}[\hspace{0.01in}\overline{\phi
(x,0)}\triangleleft\mathcal{U}^{\dag}(t,0)]\circledast\lbrack\hat
{O}\,\mathcal{U}(t,0)\triangleright\psi(\mathbf{x},0)]\nonumber\\
&  =\int\text{d}_{q}^{3}x\,\overline{\phi(\mathbf{x},0)}\circledast
\lbrack\,\mathcal{U}^{-1}(t,0)\hspace{0.01in}\hat{O}\,\mathcal{U}%
(t,0)\triangleright\psi(\mathbf{x},0)]. \label{RecHeiBil1}%
\end{align}
From the above result, we can read off an expression for observables of the
\textit{Heisenberg picture}:%
\begin{equation}
\hat{O}_{H}=\hspace{0.01in}\mathcal{U}^{-1}(t,0)\hspace{0.01in}\hat
{O}\,\mathcal{U}(t,0)=\hspace{0.01in}\mathcal{U}(0,t)\hspace{0.01in}\hat
{O}\,\mathcal{U}^{-1}(0,t). \label{ZusSchHeiBilOpeQuaDym}%
\end{equation}
Note that the second expression is a consequence of
Eq.~(\ref{UntZeitEntOpeDreQuaEuk}) in the previous chapter. A look at
Eq.~(\ref{RecHeiBil1}) also shows that the wave function of the Heisenberg
picture does not depend on time:%
\begin{equation}
\phi_{H}(\mathbf{x})=\phi(\mathbf{x},t=0).
\end{equation}
With these conventions, the Heisenberg picture yields the same matrix elements
as the Schr\"{o}dinger picture:%
\begin{align}
\big \langle\phi|\hat{O}\triangleright\psi\big \rangle_{q}  &  =\int
\text{d}_{q}^{3}x\,\overline{\phi(\mathbf{x},t)}\circledast\hat{O}%
\triangleright\psi(\mathbf{x},t)\nonumber\\
&  =\int\text{d}_{q}^{3}x\,\overline{\phi(\mathbf{x})}\circledast\hat{O}%
_{H}\triangleright\psi_{H}(\mathbf{x})=\big \langle\phi_{H}|\hspace
{0.01in}\hat{O}_{H}\triangleright\psi_{H}\big \rangle_{q}.
\label{UebEleHeiOpe1}%
\end{align}

As is well-known, the observables of the Heisenberg picture fulfill the
so-called \textit{Heisenberg equations of motion}. If the Hamilton operator
shows no explicit time-dependence, it holds in complete analogy to the
undeformed case \cite{Lavagno:2006jx}:%
\begin{align}
\frac{\text{d}\hat{O}_{H}}{\text{d}t}  &  =\frac{\partial\hspace
{0.01in}\mathcal{U}^{-1}(t,0)}{\partial\hspace{0.01in}t}\,\hat{O}%
\,\mathcal{U}(t,0)+\hspace{0.01in}\mathcal{U}^{-1}(t,0)\,\hat{O}%
\,\frac{\partial\hspace{0.01in}\mathcal{U}(t,0)}{\partial\hspace{0.01in}%
t}\nonumber\\
&  =\text{i}H\,\mathcal{U}^{-1}(t,0)\hspace{0.01in}\hat{O}\,\mathcal{U}%
(t,0)-\hspace{0.01in}\mathcal{U}^{-1}(t,0)\hspace{0.01in}\hat{O}%
\,\mathcal{U}(t,0)\,\text{i}H\nonumber\\
&  =\text{i}[H,\hat{O}_{H}],\label{HeiBewGleOpe1}\\[0.1in]
\frac{\text{d}\hat{O}_{H}}{\text{d}t}  &  =\frac{\partial\hspace
{0.01in}\mathcal{U}(0,t)}{\partial\hspace{0.01in}t}\,\hat{O}\,\mathcal{U}%
^{-1}(0,t)+\hspace{0.01in}\mathcal{U}(0,t)\,\hat{O}\,\frac{\partial
\hspace{0.01in}\mathcal{U}^{-1}(0,t)}{\partial\hspace{0.01in}t}\nonumber\\
&  =\text{i}H\,\mathcal{U}(0,t)\hspace{0.01in}\hat{O}\,\mathcal{U}%
^{-1}(0,t)-\hspace{0.01in}\mathcal{U}(0,t)\hspace{0.01in}\hat{O}%
\,\mathcal{U}^{-1}(0,t)\,\text{i}H\nonumber\\
&  =\text{i}[H,\hat{O}_{H}]. \label{HeiBewGleOpe2}%
\end{align}
Note that we have to add $\partial\hat{O}_{H}\hspace{-0.01in}/\partial t$ on
the right-hand side of the above equations of motion if $\hat{O}$ shows an
explicit time dependence.

\appendix

\section{Scalar product for the $q$-deformed Euclidean space}

In the following, we show that the expression%
\begin{equation}
\big \langle f|\hspace{0.01in}g\big \rangle_{q}=\int\text{d}_{q}%
^{3}x\,\overline{f(\mathbf{x})}\circledast g(\mathbf{x}) \label{DefSesqWdh}%
\end{equation}
has all the properties of a scalar product.

Recall that the star-product is distributive, and the $q$-integral over the
Euclidean quantum space is linear. Therefore, the expression in
Eq.~(\ref{DefSesqWdh}) is antilinear in its first argument, and it is linear
in its second argument:%
\begin{align}
\left\langle f|\hspace{0.01in}g_{1}+g_{2}\right\rangle _{q}  &  =\left\langle
f|\hspace{0.01in}g_{1}\right\rangle _{q}+\left\langle f|\hspace{0.01in}%
g_{2}\right\rangle _{q}, & \left\langle f|\alpha\hspace{0.01in}g\right\rangle
_{q}  &  =\alpha\left\langle f|\hspace{0.01in}g\right\rangle _{q},\nonumber\\
\left\langle f_{1}+f_{2}|\hspace{0.01in}g\right\rangle _{q}  &  =\left\langle
f_{1}|\hspace{0.01in}g\right\rangle _{q}+\left\langle f_{2}|\hspace
{0.01in}g\right\rangle _{q}, & \left\langle \alpha\hspace{0.01in}%
f|g\right\rangle _{q}  &  =\bar{\alpha}\left\langle f|\hspace{0.01in}%
g\right\rangle _{q}. \label{ProSes}%
\end{align}
Note that $\alpha$ is a complex number with $\bar{\alpha}$ being its complex
conjugate. Due to the identities of Eq.~(\ref{ProSes}), the expression in
Eq.~(\ref{DefSesqWdh}) is a so-called sesquilinear form.

It follows from the conjugation properties of star-product and $q$-integral
[cf. Eq.~(\ref{KonEigSteProFkt}) of\ Chap.~\ref{KapStePro} and
Eq.~(\ref{KonEigVolInt}) of Chap.~\ref{KapIntegral}] that the bilinear form in
Eq.~(\ref{DefSesqWdh}) is also conjugate-symmetrical:%
\begin{align}
\overline{\left\langle f|\hspace{0.01in}g\right\rangle }  &  =\overline
{\int\text{d}_{q}^{3}x\,\overline{f(\mathbf{x})}\circledast g(\mathbf{x}%
)}=\int\text{d}_{q}^{3}x\,\overline{\overline{f(\mathbf{x})}\circledast
g(\mathbf{x})}\nonumber\\
&  =\int\text{d}_{q}^{3}x\,\overline{g(\mathbf{x})}\circledast f(\mathbf{x}%
)=\left\langle g|f\right\rangle . \label{SymDefSes}%
\end{align}

Next, we prove that the bilinear form in Eq.~(\ref{DefSesqWdh}) is positive
definite for all $q\ $from a neighborhood of $1$ \cite{Fiore:1993os}. To this end, we first show that there can be no function
$f_{q}$ other than the zero function fulfilling the following condition:%
\begin{equation}
\big \langle f_{q}|f_{q}\big \rangle_{q}=\int_{-\infty}^{+\infty}\text{d}%
_{q}^{3}x\,\overline{f_{q}(\mathbf{x})}\circledast f_{q}(\mathbf{x})=0.
\label{PosDefQSes}%
\end{equation}
We assume that $f_{q}$ is a function subject to the above identity. Moreover,
we can assume that $f_{q}$ depends continuously on the deformation parameter
$q$. For this reason, we write $f_{q}$ as%
\begin{equation}
f_{q}(\mathbf{x})=\sum_{k\hspace{0.01in}=\hspace{0.01in}0}^{\infty}%
\hspace{0.01in}f_{q,k}(\mathbf{x})\hspace{0.01in}(q-1)^{k}
\label{EntFktPosDef}%
\end{equation}
with $f_{q,k}(\mathbf{x})$ showing the following property:%
\begin{equation}
\lim_{q\hspace{0.01in}\rightarrow1}f_{q,k}(\mathbf{x})=f_{1,k}(\mathbf{x}%
)=0\qquad\Rightarrow\qquad f_{q,k}(\mathbf{x})=0. \label{IdeEntFktNullN}%
\end{equation}
The condition in Eq.~(\ref{IdeEntFktNullN}) guarantees that $f_{q,k}$ does not
vanish for $q=1$ unless $f_{q,k}$ is identical to the zero function for any
value of $q$\ from a neighborhood of $1$.

The bilinear form in Eq.~(\ref{DefSesqWdh}) becomes the usual scalar product
in the limiting case $q=1$. For this reason, we have:%
\begin{equation}
0=\lim_{q\hspace{0.01in}\rightarrow1}\int_{-\infty}^{+\infty}\text{d}_{q}%
^{3}x\,\overline{f_{q}(\mathbf{x})}\circledast f_{q}(\mathbf{x})=\int
_{-\infty}^{+\infty}\text{d}^{3}x\,\overline{f_{1,0}(\mathbf{x})}\cdot
f_{1,0}(\mathbf{x}). \label{VerNullFktPosDef}%
\end{equation}
Since the ordinary scalar product on the right side of
Eq.~(\ref{VerNullFktPosDef}) is positive definite, it holds $f_{1,0}=0$ and,
because of Eq.~(\ref{IdeEntFktNullN}), we also have $f_{q,0}=0$. Next, we
insert the expansion of Eq.~(\ref{EntFktPosDef}) into the expression of
Eq.~(\ref{PosDefQSes}):%
\begin{align}
0  &  =\int_{-\infty}^{+\infty}\text{d}_{q}^{3}x\sum_{k,l\hspace
{0.01in}=\hspace{0.01in}1}^{\infty}\,\overline{f_{q,k}(\mathbf{x})}\circledast
f_{q,l}(\mathbf{x})\,(q-1)^{k\hspace{0.01in}+\hspace{0.01in}l}\nonumber\\
&  =(q-1)^{2}\int_{-\infty}^{+\infty}\text{d}_{q}^{3}x\,\overline
{f_{q,1}(\mathbf{x})}\circledast f_{q,1}(\mathbf{x})+O((q-1)^{3}).
\end{align}
This way, we get:%
\begin{equation}
\int_{-\infty}^{+\infty}\text{d}_{q}^{3}x\,\overline{f_{q,1}(\mathbf{x}%
)}\circledast f_{q,1}(\mathbf{x})=O(q-1).
\end{equation}
Accordingly, the integral on the left-hand side of the above equation must
vanish in the limiting case $q\rightarrow1$:%
\begin{equation}
0=\lim_{q\hspace{0.01in}\rightarrow1}\int_{-\infty}^{+\infty}\text{d}_{q}%
^{3}x\,\overline{f_{q,1}(\mathbf{x})}\circledast f_{q,1}(\mathbf{x}%
)=\int_{-\infty}^{+\infty}\text{d}^{3}x\,\overline{f_{1,1}(\mathbf{x})}\cdot
f_{1,1}(\mathbf{x}).
\end{equation}
From the positive definiteness of the ordinary scalar product follows
$f_{1,1}=0$. Due to Eq.~(\ref{IdeEntFktNullN}), this implies that $f_{q,1}$ is
also identical to the zero function. We can repeat this argument for all
$f_{q,k}$ with $k\geq2$, one after the other. Thus, we have $f_{q,k}=0$ for
all $k\in\mathbb{N}_{0}$. Therefore $f_{q}=0$ is valid. In other words, no
function different from the zero function satisfies the identity in
Eq.~(\ref{PosDefQSes}):%
\begin{equation}
\big \langle f_{q}|f_{q}\big \rangle_{q}=0\quad\Leftrightarrow\quad f_{q}=0.
\label{NicIndSesFkt}%
\end{equation}

Now we are ready to complete the proof that the bilinear form in
Eq.~(\ref{DefSesqWdh}) is positive definite for all $q$ from a neighborhood of
$1$, i.~e.%
\begin{equation}
\big \langle f|f\big \rangle_{q}>0\quad\text{for}\quad f\neq0.
\label{PosDefSesEnd}%
\end{equation}
In the following, $B(1)$ is a neighborhood of $1$, so that
Eq.~(\ref{NicIndSesFkt}) is satisfied for all $q\in B(1)$. We assume that
there is a $q^{\ast}\hspace{-0.01in}\in B(1)$ and a function $f\neq0$ such
that%
\begin{equation}
\big \langle f|f\big \rangle_{q^{\ast}}\hspace{-0.01in}<0.
\end{equation}
This assumption together with%
\begin{align}
\lim_{q\hspace{0.01in}\rightarrow1}\big \langle f|f\big \rangle_{q}  &
=\lim_{q\hspace{0.01in}\rightarrow1}\int_{-\infty}^{+\infty}\text{d}_{q}%
^{3}x\,\overline{f_{q}(\mathbf{x})}\circledast f_{q}(\mathbf{x})\nonumber\\
&  =\int_{-\infty}^{+\infty}\text{d}^{3}x\,\overline{f(\mathbf{x})}\cdot
f(\mathbf{x})=\big \langle f|f\big \rangle>0
\end{align}
implies that a $q^{\ast\ast}\hspace{-0.01in}\in B(1)$ exists with%
\begin{equation}
\big \langle f|f\big \rangle_{q^{\ast\ast}}\hspace{-0.01in}=0.
\end{equation}
However, this conclusion contradicts Eq.~(\ref{NicIndSesFkt}) due to $f\neq0$,
i.~e. Eq.~(\ref{PosDefSesEnd}) must apply to all $q\in B(1)$.

\end{document}